\documentclass[review,12pt]{elsarticle}

\usepackage{amssymb}
\usepackage{amsthm}
\usepackage{amsmath}
\usepackage{mathrsfs}
\usepackage{graphicx}
\usepackage{epstopdf}
\usepackage{float}
\usepackage{caption}
\usepackage{subcaption}
\usepackage{makecell}
\usepackage{bm}
\usepackage{bbm}
\usepackage{booktabs}
\usepackage{mathrsfs}
\usepackage{multirow}
\usepackage{parskip}
\usepackage{amsmath}
\usepackage{fancyhdr}
\usepackage{chngcntr}
\usepackage{apptools}
\usepackage{mathrsfs}
\usepackage{hyperref}
\usepackage{cleveref}
\usepackage{soul}
\usepackage{color,soul} 
\usepackage{color} 
\usepackage{siunitx}
\usepackage[margin=2.5cm]{geometry}
\usepackage[table]{xcolor}
\biboptions{sort&compress}
\graphicspath{ {./images/} }

\journal{Elsevier}

\makeatletter
\def\@author#1{\g@addto@macro\elsauthors{\normalsize%
    \def\baselinestretch{1}%
    \upshape\authorsep#1\unskip\textsuperscript{%
      \ifx\@fnmark\@empty\else\unskip\sep\@fnmark\let\sep=,\fi
      \ifx\@corref\@empty\else\unskip\sep\@corref\let\sep=,\fi
      }%
    \def\authorsep{\unskip,\space}%
    \global\let\@fnmark\@empty
    \global\let\@corref\@empty  
    \global\let\sep\@empty}%
    \@eadauthor={#1}
}
\makeatother

\begin{document}

\begin{frontmatter}



\title{Inverse Design of Cellular Composites for Targeted Nonlinear Mechanical Response via Multi-Fidelity Bayesian Optimisation}



\author{Hirak Kansara\fnref{QMUL}}
\author{Leo Guo\fnref{TU}}

\author{Wei Tan\corref{cor1}\fnref{QMUL}} \ead{wei.tan@qmul.ac.uk}

\address[QMUL]{School of Engineering and Materials Science, Queen Mary University of London, Mile End Road, London, E1 4NS, UK}

\address[TU]{Department of Materials Science and Engineering, Delft University of Technology, 2628 CD Delft}

\cortext[cor1]{Corresponding author.}

\begin{abstract}

The rise of machine learning and additive manufacturing has enabled the design of architected materials with tailored properties that surpass those of natural materials. Inverse design offers a data-efficient alternative to trial-and-error methods, yet most existing approaches depend on either large datasets or scarce high-fidelity data from simulations and experiments. These requirements pose a particular challenge for architected materials with nonlinear mechanical responses, where capturing complex deformation modes requires expensive evaluations. To address this, a Multi-Fidelity Bayesian Optimisation (MFBO) framework for the inverse design of cellular composites that directly targets their full nonlinear response is introduced. By integrating information from multiple fidelity sources and scalarising the response using a similarity score, the framework enables efficient exploration of the design space while reducing reliance on costly evaluations. As a proof of concept, the method is applied to spinodoid cellular composites using finite element models, validated with compression tests on short carbon-fibre reinforced PET-G composites. Four target responses were considered, with three multi-fidelity strategies benchmarked against a standard single-fidelity approach. Across all cases, MFBO achieved higher similarity scores and consistently recovered the targeted responses, outperforming the single-fidelity baseline under the same evaluation budget, while also successfully recovering all targeted responses. These results demonstrate the effectiveness of MFBO for inverse design of stochastic architected materials, where high-quality data is scarce but lower-cost proxies exist. By efficiently navigating complex design spaces, MFBO enables the creation of cellular composites with precisely tailored nonlinear mechanical behaviour.

\end{abstract}


\begin{keyword}

Inverse Design \sep Bayesian Optimisation \sep Cellular Composites \sep Multi-Fidelity 



\end{keyword}

\end{frontmatter}


\section{Introduction}
\label{Introduction}

Cellular composites are advanced lightweight structures composed of interconnected networks of struts or walls, forming architectures that may be either periodic or stochastic. Enabled by recent advances in additive manufacturing, these structures can be realised using a wide range of constituent materials, including fibre-reinforced composites, resulting in an expansive design space defined by both geometry and material combinations. This versatility gives rise to a rich spectrum of mechanical behaviours, which can be tailored to meet specific performance requirements. However, practical design objectives, such as stiffness, strength, or energy absorption, constrain the set of viable configurations, raising a fundamental challenge: how to systematically identify the appropriate combination of geometry and material that yields a desired mechanical response. Traditional approaches often rely on iterative, intuition-driven exploration of the structure-property relationship, which becomes increasingly impractical as the dimensionality or nonlinearity of the design space grows. This challenge motivates the adoption of more systematic design paradigms, as discussed herein.

From a broader perspective, forward design focuses on predicting the effects that result from a specified model configuration, following a straightforward cause–and–effect relationship in which governing equations and input parameters are known. In contrast, inverse design works backwards from a desired effect to determine the underlying design parameters or model configuration required to achieve it. By directly targeting performance objectives rather than exhaustively exploring the design space, inverse approaches offer a more efficient and systematic way to identify optimal solutions. While forward problems are well-suited to analysis, they become inefficient for design exploration in high-dimensional spaces, particularly when complex nonlinear responses are involved. In such cases, inverse design provides a more direct and systematic framework for identifying structures that achieve target performance, making it a powerful paradigm for navigating complex architected material design spaces. Here, inverse design can therefore be interpreted as a specific instance of a broader class of inverse problems, in which the objective is to obtain system configurations from observed or desired responses.

Motivated by these advantages, researchers have increasingly turned to inverse design methodologies. Rather than mapping the structure-to-property relationship, inverse design focuses on directly discovering structures that exhibit desired target properties, thereby minimising reliance on expert domain knowledge \cite{kumar2020inverse, bastek2023inverse, ha2023rapid, zheng2023deepb}. While effective approaches exist, many implementations remain confined to simplified assumptions, such as linear elastic behaviour, which may fail to capture complex real-world responses. Efforts to extend these methods to account for large deformation and nonlinear material behaviour have been reported \cite{clausen2015topology, zeng2023inverse, wang2014design}, but significant challenges remain, particularly in handling complex geometries. Incorporating such nonlinearities often results in considerable computational cost, motivating the adoption of data-driven surrogate models to efficiently approximate structural responses and navigate the design space \cite{zheng2021data, bastek2022inverting, woldseth2022use, kollmann2020deep}.

An alternative strategy involves directly mapping performance metrics to design parameters \cite{zheng2023deep}.The inherent characteristic of such strategies is that multiple distinct design configurations can correspond to the same set of performance metrics, which leads to a many-to-one relationship. This non-uniqueness makes the inverse design mapping ill-posed, as it lacks a well-defined solution. Approaches such as dual-network models have been proposed to achieve unique mappings \cite{liu2018training, kumar2020inverse, ha2023rapid}, although they may limit generalisation and restrict the ability to generate multiple valid solutions. To address this, probabilistic methods have been introduced, modelling the solution space as a distribution and enabling the generation of diverse candidate designs \cite{ma2019probabilistic, zeng2023deep, bastek2022inverting}. Despite their flexibility, such methods typically require large volumes of training data, particularly when high-fidelity data are involved.

However, in many practical applications, generating such extensive datasets is prohibitively expensive. This limitation has driven the development of indirect inverse design methods, which avoid learning an explicit property-structure mapping and instead search for optimal designs through optimisation heuristics. Metaheuristic algorithms, including genetic algorithms \cite{mitchell1998introduction, wang2022inverse, deng2022inverse}, and evolutionary strategies \cite{back1996evolutionary}, have been widely applied, alongside experimentally informed approaches \cite{thakolkaran2024experimentinformedfinitestraininversedesign}. Nevertheless, these methods often suffer from inefficient sampling and sensitivity to hyperparameter tuning, limiting their effectiveness in high-cost evaluation settings. More recent approaches, such as deep neural operators \cite{jin2025characterization} and physics-informed neural networks \cite{thakolkaran2024experimentinformedfinitestraininversedesign}, offer improved efficiency but still rely on access to high-quality data.

Within this landscape, Bayesian optimisation (BO) has emerged as a sample-efficient framework for inverse design problems involving expensive evaluations \cite{brochu2010tutorial, rasmussen2003gaussian, bessa2019bayesian, frazier2018tutorial, shahriari2015taking}. Its application to architected materials has demonstrated a reduction in the number of required evaluations compared to conventional approaches \cite{rassloff2025inverse}. However, most existing implementations assume access to a single high-fidelity information source. To overcome this, multi-fidelity Bayesian optimisation (MFBO) methods have gained increasing attention \cite{takeno2020multi, wu2020practical, foumani2023multi, lee2023machine}, enabling the integration of multiple information sources, where lower-fidelity models serve as computationally inexpensive approximations of higher-fidelity simulations.

Despite the promise of MFBO, its application to inverse design frameworks remains largely focused on optimising reduced or scalar performance metrics, such as stiffness or energy absorption \cite{grbvcic2025inverse, guo2026multi}. While such quantities are convenient for optimisation, they provide only limited insight into the full mechanical behaviour of architected materials, particularly under complex loading conditions. In many practical applications, however, performance is governed not by a single metric but by the entire nonlinear stress–strain response, which captures key phenomena such as yielding, strain hardening, and failure. Although previous studies have successfully addressed multi-objective trade-offs between competing scalar metrics \cite{Kansara:MOBO(2025), satpati2025multi} and have explored inverse design targeting stress–strain curves \cite{zhang2025inverse}, these approaches generally do not incorporate multiple fidelity sources and thus remain limited in their ability to optimise the full mechanical response with very expensive problems directly. Two complementary approaches exist in the literature: multi-fidelity optimisation, which improves sample efficiency, and full-response inverse design, which captures complex mechanical behaviour, yet they remain largely disconnected. This underscores the need for an inverse design framework that simultaneously leverages multi-fidelity information while directly targeting high-dimensional stress–strain responses in a computationally efficient manner.

To directly target full stress–strain behaviour, a pointwise evaluation strategy can be employed. In this approach, stress values are sampled at fixed, uniformly spaced strain intervals across all designs, ensuring that each curve is represented on a common basis and allowing direct comparison of stress values regardless of the underlying design parameters. Despite this standardisation, the resulting curves remain inherently high-dimensional, as they consist of stress values evaluated over many strain points. This high dimensionality poses significant challenges for optimisation, particularly within BO frameworks. Several strategies have been proposed to address this issue, including principal component analysis (PCA), which projects responses into a lower-dimensional subspace \cite{raponi2020high}, autoencoders and variational autoencoders (VAEs), which learn compact latent representations \cite{griffiths2020constrained}, and scalarisation methods, which reduce multi-dimensional outputs to a single objective. While latent-variable approaches offer advantages such as transfer learning, they introduce reconstruction errors and additional modelling complexity. In contrast, scalarisation provides a practical and efficient alternative, as it does not require additional training data or constraints on explained variance.

Building on these considerations, this work presents a novel inverse design framework that integrates MFBO with scalarised objectives to efficiently target the full nonlinear mechanical response of spinodoid cellular composites. The framework addresses two key gaps: the limited integration of MFBO within black-box inverse design methodologies for high-dimensional problems, and the lack of sample-efficient strategies for matching complete mechanical response profiles under realistic loading and manufacturing conditions. The proposed approach leverages multiple fidelity sources to improve sampling efficiency and guide the optimisation process towards high-quality designs. As a proof of concept, the framework is applied to the inverse design of spinodoid structures using validated finite element models that incorporate elastic–plastic behaviour, damage evolution, manufacturing-induced anisotropy, and fabrication defects. Several acquisition functions, including both utility-based and information-theoretic strategies, are benchmarked to evaluate convergence performance relative to single-fidelity approaches, demonstrating the generality and robustness of the proposed framework.

\section{Methodology}
\label{section:methodology}

\begin{figure}[!h]
    \centering
    \includegraphics[width=1\textwidth]{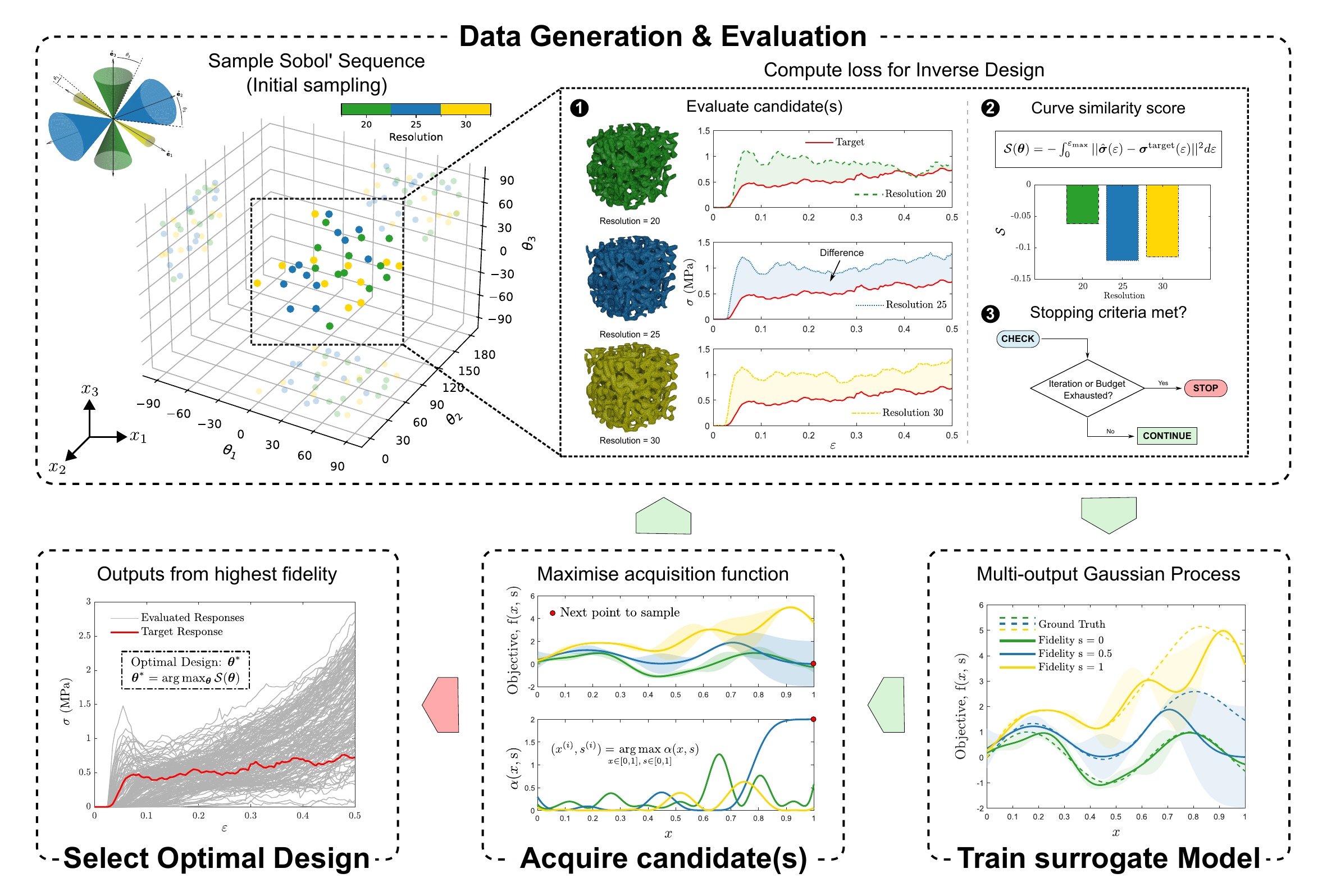}
    \caption[Workflow for inverse design using Multi-Fidelity Bayesian Optimisation.]{Workflow for inverse design using multi-fidelity Bayesian optimisation. The process begins by generating an initial pool of candidate designs using a Sobol' sequence, with each point evaluated at three fidelity levels. A similarity score quantifies the difference between simulated and target responses and serves as the optimisation objective, with inputs $\boldsymbol{\theta}$. A multi-output Gaussian process surrogate model is trained on the collected data to predict the objective behaviour across the design space. An acquisition strategy is then employed to select new candidates for evaluation (in this figure, the acquisition function aims to reduce variance per unit cost). This iterative process continues until the convergence criteria are met. Finally, the optimal design is selected at the highest fidelity, corresponding to the highest similarity value.}
    \label{fig:inverse_des_framework}
\end{figure}

This section presents the multi-fidelity inverse design framework developed to optimise the nonlinear mechanical response of spinodoid cellular composites. The proposed approach integrates finite element simulations, data-driven surrogate modelling based on multi-output GPs, and BO to identify the set of design parameters that achieve the target mechanical performance. The methodology comprises three main components: (1) formulation of the inverse design problem, including the parameterisation of spinodoid architectures and experimental validation of finite element setup to ensure predictive accuracy; (2) construction of a multi-fidelity surrogate model using multi-output GPs to capture structure-property relationships across fidelities as well as cross-fidelity knowledge transfer; and (3) implementation of various acquisition functions to sequentially and efficiently select new sample points at appropriate fidelity levels to evaluate. Ultimately, the dataset then expands iteratively throughout the optimisation process, minimising the number of expensive high-fidelity evaluations required compared to conventional BO. This workflow is depicted in Fig.~\ref{fig:inverse_des_framework}.

\subsection{Problem formulation}
\label{subsection:problem formulation}
The goal of this work is the inverse design of spinodoid cellular composites, whose topologies are generated via a Gaussian Random Field (GRF) model inspired by spinodal decomposition \cite{kumar2020inverse}. The geometry is controlled through a set of design parameters described by:

\begin{equation}
    \boldsymbol{\Theta} = \{\rho, \lambda, \theta_1, \theta_2, \theta_3\}
\end{equation}

where $\rho$ is the relative density determining porosity, $\lambda$ is the wavenumber which dictates the microstructural feature size, and $\theta_{i}^{i = 1,2,3}$ are conical angles defining anisotropy in the principal directions. These parameters together define the material's 3D architecture, allowing the properties to be tuned, resulting in isotropic to highly anisotropic topologies. However, by considering  $\rho = 0.3$ and $ \lambda = 15\pi $ to be fixed constants, this reduces the design space to a subset
\begin{equation}
    \boldsymbol{\theta} = \{\theta_1, \theta_2, \theta_3\} \subset \boldsymbol{\Theta}
\end{equation}
which defines the effective (or reduced) set of design variables. A value of 0.3 was chosen mainly to reduce computational cost, as higher values would require more elements to discretise the denser geometry, thereby increasing the simulation time. The wavenumber was selected to ensure a clear separation of scales between the microscale and macroscale, which is primarily used to obtain homogenised properties \cite{deng2024ai}.

While the full set of design parameters in $ \boldsymbol{\Theta} $ can be considered in principle, allowing for a broader range of microstructural configurations and potentially more diverse stress–strain responses, this would significantly increase the complexity of the design space. A higher-dimensional space introduces more possible parameter combinations, each potentially yielding distinct mechanical behaviour, which may increase the optimisation time. In this work, $\rho$ and $\lambda$ are fixed to demonstrate the proposed inverse design framework in a reduced setting, which serves as a proof of concept without loss of generality. In addition, FEM setup and experimental validation can be found in Supplementary material \ref{appendix:FEM setup and experimental validation}.

\begin{figure}[!h]
    \centering
    \includegraphics[width=1\textwidth]{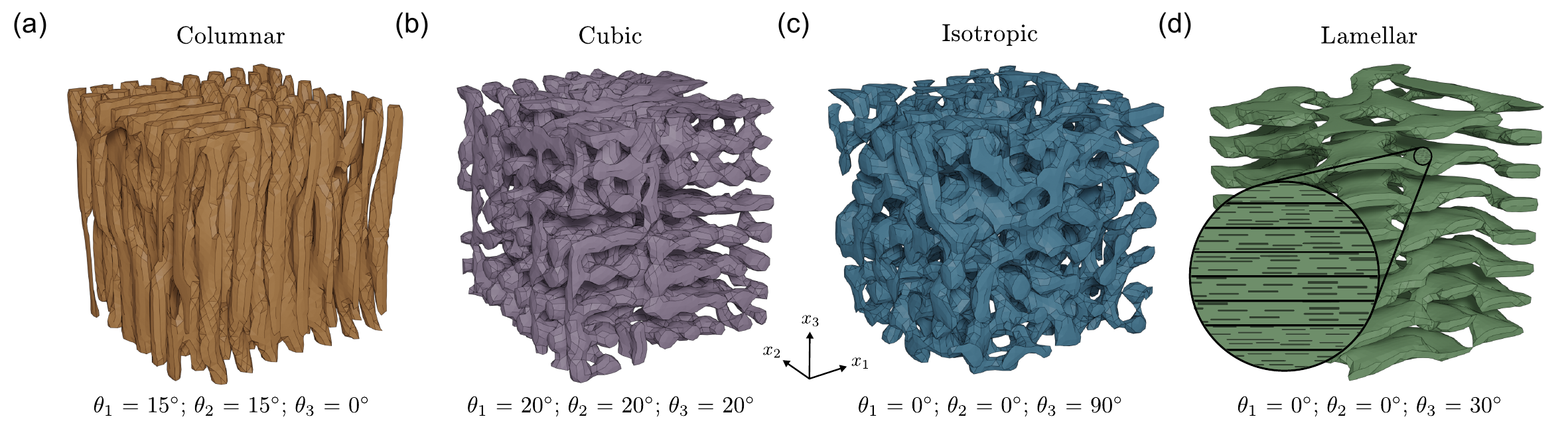}
    \caption{Four topologies (a) Columnar, (b) Isotropic, (c) Cubic, (d) Lamellar were generated by varying $\theta$ with $\rho = 0.3$ and $\lambda = 15\pi$, serving as validation benchmarks. They range from anisotropic column-like structures and layered structures to isotropic or cubic structures defined by conical angles. Schematic in (d) shows the magnified microstructure post fabrication, where short-fibres are oriented parallel to the $x_1-x_2$ plane.}
    \label{fig:spinodoids}
\end{figure}

\subsection{{Target responses}}
\label{subsection:target responses}
\begin{figure}[]
    \centering
    \includegraphics[width=1\textwidth]{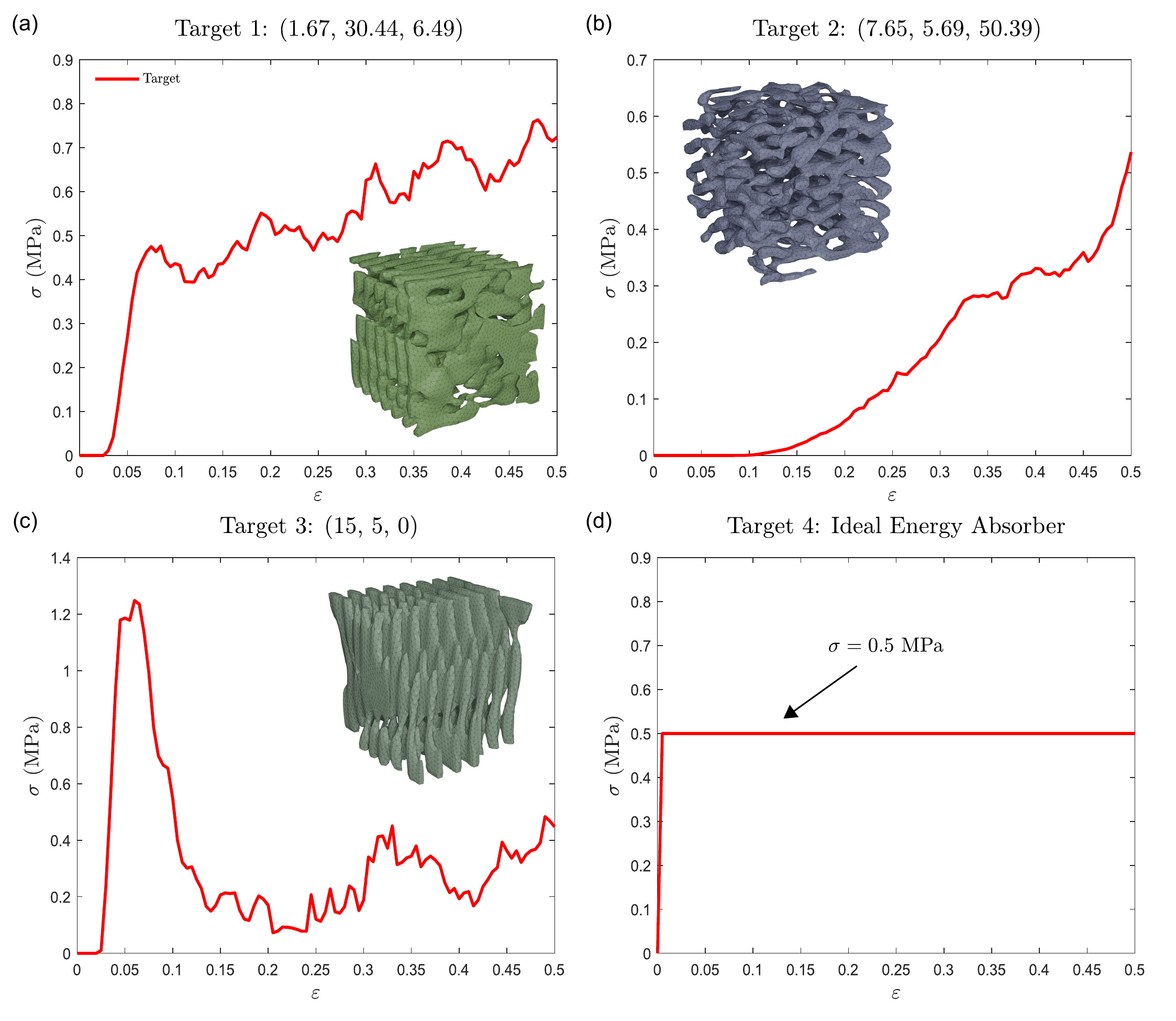}
    \caption{{Four target mechanical responses with insets of (a)-(c) showing generated structures, while (d) is an arbitrary mechanical response based on an ideal energy absorber.}}
    \label{fig:target_params}
\end{figure}

Four target cases were selected to evaluate the robustness of the inverse design framework. These targets exhibit distinct nonlinear mechanical responses and were chosen to cover both achievable designs within the design space and an idealised, energy-absorbing structure \cite{gibson2003cellular}. The generated structures, along with the mechanical responses, are illustrated in Fig.~\ref{fig:target_params}. 

\subsection{Adapting MFBO for inverse design}

Under standard optimisation settings, the goal is typically to find inputs $\textbf{x}$ that maximise or minimise an unknown objective function. In contrast, inverse design reframes the problem. Rather than searching for an extremum, the goal is to identify a set of inputs $\textbf{x}$ that produce an output as close as possible to a specified target. 


The fundamental shift requires modifying the objective function used in BO. Instead of treating the output of the black-box function as an unknown quantity to be maximised/minimised, the optimisation is reoriented towards minimising the discrepancy/similarity between the predicted output and a predefined target. Formally, if $f(\textbf{x})$ is the surrogate model prediction and $\text{y}^*$ is the target output, the inverse design problem becomes
\begin{equation}
    \textbf{x}_{\text{inv}} = \arg\min_{\textbf{x}} ||f(\textbf{x})-\text{y}^*||^2 = \arg\max_{\textbf{x}} -||f(\textbf{x})-\text{y}^*||^2
\end{equation}
where $\textbf{x}_{\text{inv}}$ is the input value that minimises the discrepancy between the predicted and the target. This frames the optimisation problem as one of seeking optima to similarity matching.

This work adopts a scalarisation approach for simplicity and interpretability, as well as ease of implementation. The scalarised objective function is defined by a similarity score $\mathcal{S}$, where
\begin{equation}
\mathcal{S}(\boldsymbol{\theta})  = -\int^{\varepsilon_{\max}}_0||\hat{\boldsymbol{\sigma}}(\varepsilon)-\boldsymbol{\sigma}^{\text{target}}(\varepsilon)||^2 d\varepsilon \approx - \sum_{i=1}^{N}(\hat{\sigma}_i-\sigma^{\text{target}}_i)^2
\end{equation}
Here, $\hat{\sigma}_i = \hat{\sigma}(\varepsilon_i)$ and $\sigma^{\text{target}}_i = \sigma^{\text{target}}(\varepsilon_i)$ correspond to stress values at the $i^{\text{th}}$ uniformly spaced strain point, with $N = 101$ points sampled between 0 and $\varepsilon_{\max} = 0.5$.

\subsection{{Mesh resolution as different fidelities}}

\begin{figure}[!h]
    \centering
    \includegraphics[width=1\textwidth]{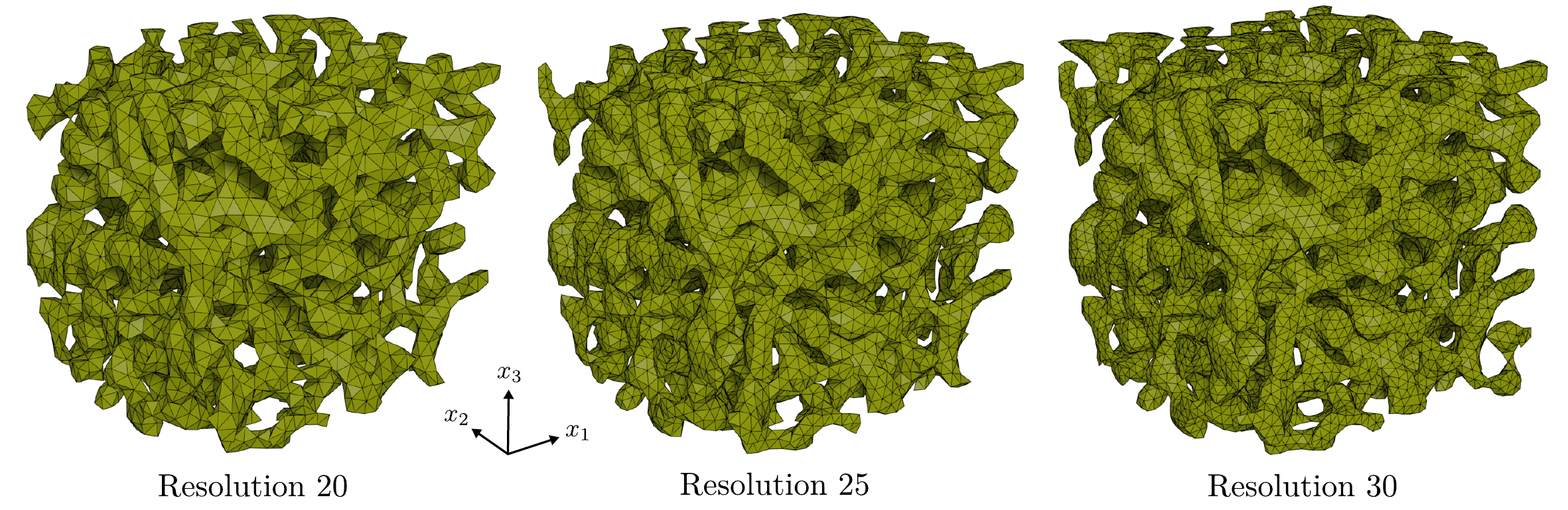}
    \caption[Illustration of how mesh resolution influences the generated structure, using the `Isotropic' topology as an example.]{{Illustration of how mesh resolution influences the generated structure, using the `Isotropic' topology as an example. Resolution increases from 20 to 30 (left to right). This primarily changes the number of elements, whereby resolution of 20 results in 13,562 elements, 22,622 elements for resolution of 25, and 33,634 for resolution of 30.}}
    \label{fig:effect_of_resolution}
\end{figure}

{Evaluating $\mathcal{S}(\boldsymbol{\theta})$ remains computationally expensive due to the cost of FEM simulations, whose runtime is strongly influenced by mesh resolutions. Finer meshes reduce discretisation error and better capture mechanical behaviour, but at substantially higher computational cost. Coarser meshes, by contrast, are far cheaper to evaluate but introduce numerical error and increased output variance, reducing surrogate model reliability. Fig.~\ref{fig:effect_of_resolution} illustrates how resolution affects the generated geometry and mesh density.}

{To balance accuracy and efficiency, a mesh convergence study was first conducted to identify a suitable reference resolution. However, for nonlinear simulation with complex geometries, even this converged mesh is too costly to apply uniformly across the entire design space. This motivates the use of a multi-fidelity framework, where simulations at different mesh resolutions, and therefore different numerical fidelities, are combined to accelerate optimisation.}

To assess the suitability of using multiple mesh resolutions as fidelity levels, the correlation between outputs obtained at different resolutions was evaluated using an initial Sobol' sample of 35 design points. Strong Pearson correlation coefficients were observed between the medium- and high-resolution simulations, with slightly lower, but still informative, correlations for the coarse mesh. These results confirm that effective information transfer across fidelities is feasible (complete results provided in Supplementary material \ref{appendix:Pearson correlation coefficients for Targets 2, 3, and 4}).

\subsection{Multi-fidelity Bayesian optimisation}
\label{subsection: Multi-fidelity Bayesian Optimisation}
Real-world optimisation problems, including the one addressed in this study, typically involve the optimisation of an objective function, $f$. Evaluating this function can be computationally expensive, especially when using high-fidelity simulations such as fine-mesh FEM models or detailed physical experiments. However, in some cases, it is possible to obtain lower-fidelity approximations of the objective functions. These approximations may be less accurate, but can still provide valuable information about the underlying function landscape.

To take advantage of this structure, MFBO integrates information from multiple fidelity levels to accelerate convergence. Each fidelity level is treated as a different but related task, denoted as $f^{(s)}(\mathbf{x})$, where $s \in \{1, \dots, S\}$ indexes the fidelity level. These are ordered such that $s = 1$ represent the lowest fidelity and $s = S $ the highest (most accurate yet expensive) fidelity. It should be noted that here, it is assumed that fidelity levels are discrete. However, this strategy can be extended to a continuous fidelity space \cite{kandasamy2017multi}.

\subsubsection{Multi-output Gaussian processes}
BO relies on a probabilistic surrogate model to guide the search for the optimum. For this, GPs are commonly used as surrogate models due to their ability to provide predictions for the mean and uncertainty \cite{rasmussen2003gaussian}. These are two common ways to incorporate multiple fidelities into GP surrogate models, which are then used within BO:

(1) Single-output GP

A straightforward approach is to treat all outputs as coming from a standard single-output GP (Eq.~\ref{equ:single-task GP}), where the fidelity level $s$ is considered as another input parameter. In this formulation, the GP learns a single posterior over all the joint input-fidelity space. This approach is simple to implement and imposes a lower computational burden compared to more complex GP-based surrogates. Moreover, it also provides an interpretable mapping from inputs to outputs. However, a significant caveat of this approach is the lack of inter-fidelity knowledge transfer, which limits data efficiency. Nonetheless, this method can still be effective in scenarios where outputs across fidelities are not strictly ordered, for example, when lower-fidelity evaluations do not consistently fall below those at the highest fidelity \cite{irshad2024leveraging}.

The standard single-output GP prior is defined as
\begin{equation}
    f(\mathbf{x}) \sim \mathcal{GP}(m(\mathbf{x}), k(\mathbf{x},\mathbf{x}'))
    \label{equ:single-task GP}
\end{equation}

where $m(\mathbf{x})$ is the mean function, and $k(\mathbf{x},\mathbf{x}')$ is the covariance kernel (e.g., RBF, Matern, etc). Also, recalling the predictive distribution for test input $\textbf{x}^*$ and training set $O = \{\mathbf{X} ,\textbf{y}\} = \{ (\textbf{x}_i,\text{y}_i) \}^N_{i=1}$ is obtained using
\begin{align}
    \mu(\mathbf{x}^*) &= k(\mathbf{X}, \mathbf{x}^*)^\top K_X^{-1} \mathbf{y} \\
    \sigma^2_{\text{post}}(\mathbf{x}^*) &= k(\mathbf{x}^*, \mathbf{x}^*) - k(\mathbf{X}, \mathbf{x}^*)^\top K_X^{-1} k(\mathbf{X}, \mathbf{x}^*)
\end{align}

where \textbf{X} are the training points and $K_X = k(\mathbf{X}, \mathbf{X}) + \sigma_n^2 \mathbf{I}$ and $\sigma_n^2$ denotes the observation noise variance.

(2) Multi-task Gaussian process

A multi-output GP generalises the standard single-output GP by jointly modelling multiple related outputs and capturing the correlations between fidelity levels, which is especially beneficial in data-scarce multi-fidelity settings. Let $f^{s}(\mathbf{x})$ denote the function at fidelity level $s$. The joint prior is 
\begin{equation}
    {\mathbf{f}(\textbf{x}) \sim \mathcal{GP}(\textbf{m}(\textbf{x}),\mathcal{K}(\textbf{x},\textbf{x}'))}
\end{equation}
where $\mathbf{f}(\textbf{x}) = [f^{(1)}(\textbf{x}),...,f^{(S)}(\textbf{x})]^\top$, and $\textbf{m}(\textbf{x}) = [m^{(1)}(\textbf{x}),...,m^{(S)}(\textbf{x})]^\top$, typically assumed to be zero, with outputs being normalised and standardised and $\mathcal{K}$ encodes both input-space and cross-fidelity correlations.
The Intrinsic Coregionalisation Model (ICM) is adopted, in which the covariance between fidelity levels $s$ and $s'$ factorises as
\begin{equation}
    {k_{s,s'}(\textbf{x}, \textbf{x}') = \textbf{B}_{s,s'}k(\textbf{x}, \textbf{x}')}
\end{equation}
{where $\textbf{B}\in \mathbb{R}^{S \times S}$ is a positive semi-definite coregionalisation matrix. This induces a block-structured covariance matrix over all observed data.}

{Similar to a single-output GP, given an observed dataset $(\textbf{X}, \textbf{y})$, to obtain the predictive posterior for a test point $\textbf{x}^*$ and $s^*$, the following equations can be utilised}
\begin{align}
    \mu^{s^*}(\mathbf{x}^*) &= \textbf{k}_*^\top \textbf{K}_{\text{MT}}^{-1} \mathbf{y} \\
    \sigma^{2s^*}_{\text{post}}(\mathbf{x}^*) &=\textbf{k}_{**} - \textbf{k}_*^\top\textbf{K}_{\text{MT}}^{-1} \textbf{k}_*
\end{align}
{where $\textbf{k}_* = [k_{s^*,s}(\textbf{x}^*,\textbf{x}_i^s)]_{s=1,...,S;i=1,...,N_s}$ is the covariance vector between the test point and training points across all fidelity levels and $\textbf{k}_{**}=k_{s^*,s^*}(\textbf{x}^*,\textbf{x}^*)$ is the variance at the test point. This results in a similar structure as the single-fidelity case, but the covariances are vectorised  (full details in Supplementary material \ref{appendix: Further MTGP details and other multi-output GPs})}. In this study, the mesh resolution is varied to obtain evaluations at different fidelity levels. The physical interpretation of mesh resolution is demonstrated in the succeeding sections.

\subsubsection{Multi-fidelity acquisition functions}
\label{subsubsection:Multi-fidelity acquisition functions}

The primary goal of MFBO is to leverage inexpensive low-fidelity approximations to improve the surrogate model's understanding of the objective at the highest fidelity. While the MTGP surrogate model captures the relationship between the design variables and outputs across fidelities, an effective strategy is still required to navigate this surrogate space. This is achieved through acquisition functions, which jointly determine both the next input location and the fidelity level for the next evaluation. 

{Most MFBO acquisition functions extend single-fidelity formulations by incorporating fidelity-dependent costs and correlations. In this work, three representative examples are considered: Multi-Fidelity Expected Improvement (MF-EI), Multi-Fidelity Upper Confidence Bound (MF-UCB), and sequential Multi-Fidelity Max Entropy Search (MF-MES) strategy using Logarithmic Expected Improvement (Log-EI). A brief overview is given below,

{(1) \emph{MF-EI}}

{This acquisition function is built on top of the single-fidelity Log-EI function, which is defined as}
{\begin{equation}
    \text{Log-EI}(\textbf{x}) = \text{Log}\left[ \text{EI}(\textbf{x}) + \epsilon \right]
\end{equation}}
{where EI depends on the GP posterior mean and variance. In MFBO, Log-EI is scaled by (i) the \emph{cost ratio} (CR) between fidelities and (ii) the correlation ($\tilde{\rho}$) between low- and high-fidelity outputs. This yields}
{\begin{equation}
    \text{MF-Log-EI} = \text{Log-EI}(\textbf{x}) \cdot \text{CR} \cdot \tilde{\rho}(s)
\end{equation}}
{where the correlation term encourages using fidelities that better approximate the highest fidelity.}

{(2) \emph{MF-UCB}}

{In contrast, MF-UCB does not rely on such multiplicative fidelity correlations. Instead, it retains the UCB structure but uses MTGP predictions at each fidelity}
{\begin{equation}
    \text{MF-UCB}(\textbf{x},s) = \omega_1 \cdot \mu^s(\textbf{x})+\omega_2 \cdot \sigma^s_{\text{post}}(\textbf{x}) \cdot \text{CR}
\end{equation}}
{where the adaptive weights $\omega_1$, $\omega_2$, are derived from the highest-fidelity posterior and balance exploitation and exploration. As only the uncertainty term is cost-scaled, the method prefers exploratory sampling at lower fidelities while exploiting high-fidelity predictions only when uncertainty is low.}

{(3) \emph{Sequential MF-MES}}

{A two-step approach is utilised where the sampling decision is made sequentially. First, the next location is found by using the predictions at the highest fidelity, which is undertaken using the Log-EI acquisition function,}
{\begin{equation}
    \textbf{x}_{\text{next}} = \arg\max \text{Log-EI}(\textbf{x}, S)
    \label{equ:log EI at high fid}
\end{equation}}
{followed by the choice of fidelity at which to evaluate, found by maximising the information gain per unit cost,}
{\begin{equation}
    s_{\text{next}} = \frac{I(\text{y}^*;\text{y}^s|O,\textbf{x}_{\text{next}})}{C(s)}.
\end{equation}}
Full expressions for the mutual-information term and its multi-fidelity extension are given in Supplementary material \ref{appendix:Detailed derivations and multi-fidelity acquisition formulations}, along with the full Mathematical formulations.

 It should be noted that the kernel functions used in GP regression require tuning the model parameters, which are additional parameters within the kernel that control properties such as smoothness, length-scale, signal variances, etc. For clarity, the notation for these parameters has been suppressed in the descriptions of single-task GPs and MTGPs. Hyperparameter learning is a crucial part of GP training and typically occurs whenever the surrogate model is updated. In practice, this is achieved by maximising the (marginal) log-likelihood (MLL) using multi-start, gradient-based optimisers such as Adam or L-BFGS-B. In addition, the acquisition function optimisation requires solving a separate inner-loop optimisation problem to identify the next evaluation point. This study utilised the L-BFGS-B algorithm for optimising both the MLL and the acquisition function with 10 random restarts, initialised from 512 quasi-random samples of the design space. These settings help mitigate poor local optima and improve the robustness of the optimisation process.

{Although hyperparameter tuning and acquisition function optimisation are central to building accurate GP surrogates, the overall efficiency of the optimisation framework also depends on the computational cost of evaluating candidate designs. To characterise this, simulation wall times were measured across representative spinodoid structures at different mesh resolutions. Because evaluation time varies with both resolution and design parameters $\boldsymbol{\theta}$, a dedicated GP surrogate was trained to model this cost, enabling fidelity-aware acquisition. Additional details on the MTGP and BO methodology, including budget management, stopping criteria, surrogate model choices, modelling assumptions, and computational configuration, are provided in Supplementary \ref{appendix:Additional details of MTGP and BO methodology}}.

\section{Results}
\label{sec:results MFBO}

{Four targets were utilised to assess the performance of the proposed framework, as outlined in Subsec.~\ref{subsection:target responses}. Across all targets, the same initial dataset was employed. While the choice of initial dataset can influence the rate of convergence—particularly if the Sobol' sequence yields samples that are locally concentrated in objective space due to nonlinear input–output mappings—it provides a consistent and unbiased starting point for comparison. This ensures a fair evaluation across different target cases and allows the robustness of the inverse design framework to be assessed under uniform initial conditions. Benchmark studies on synthetic multi-fidelity functions further demonstrate that optimisation performance can be sensitive to the specific Sobol' realisation used to generate the initial dataset, especially in higher-dimensional settings, leading to increased run-to-run variability (Further details on benchmarking can be found in Supplementary information \ref{appendix:Synthetic function benchmarking}). Fixing the initial dataset therefore avoids introducing additional sources of variance and ensures that observed performance differences arise from the optimisation strategies themselves rather than from stochastic initialisation effects. This effect is further mitigated by the reduced dimensionality of the spinodoid design space. While reusing data from previous inverse design runs would be both feasible and practical in real-world applications, potentially improving efficiency, it is omitted here to avoid introducing unnecessary bias.}

{The optimisation begins by using a Sobol' sequence to sample the input parameter space $\boldsymbol{\theta}$. Depending on whether a single- or multi-fidelity strategy is employed, the initial dataset consists of evaluations performed either exclusively at the highest fidelity or distributed across multiple fidelity levels. In all cases, the total computational cost of the initial dataset is fixed at 30,000 seconds to ensure a fair comparison between methods. Four optimisation strategies are considered, as outlined in Subsec.~\ref{subsubsection:Multi-fidelity acquisition functions}. }

{The resulting dataset is used to train a GP surrogate model, which provides a probabilistic approximation of the objective function over the design space. BO then proceeds iteratively by using this surrogate to predict both the mean and uncertainty of the similarity score of unobserved inputs. At each iteration, an acquisition function leverages this information to balance exploration and exploitation of already evaluated regions of the design space, selecting the next input parameters and in the multi-fidelity case, the fidelity level to be evaluated. The newly obtained evaluation is subsequently incorporated into the dataset, and the GP model is updated, allowing the optimisation to progressively refine its estimate of the optimal design parameters. Further details can be found in \cite{rasmussen2006gaussian}.}

\subsection{{Optimisation walk-through of Target 1 (Illustrative Example)}}

{To illustrate the operation of the proposed Bayesian inverse design framework, Target 1 is examined in detail as a representative example. This case is used to demonstrate how different acquisition strategies interact with different fidelity levels, how candidate optima are identified, and how convergence is achieved under a fixed computational budget for both single- and multi-fidelity methods.}

\begin{figure}[]
    \centering
    \includegraphics[width=0.9\textwidth]{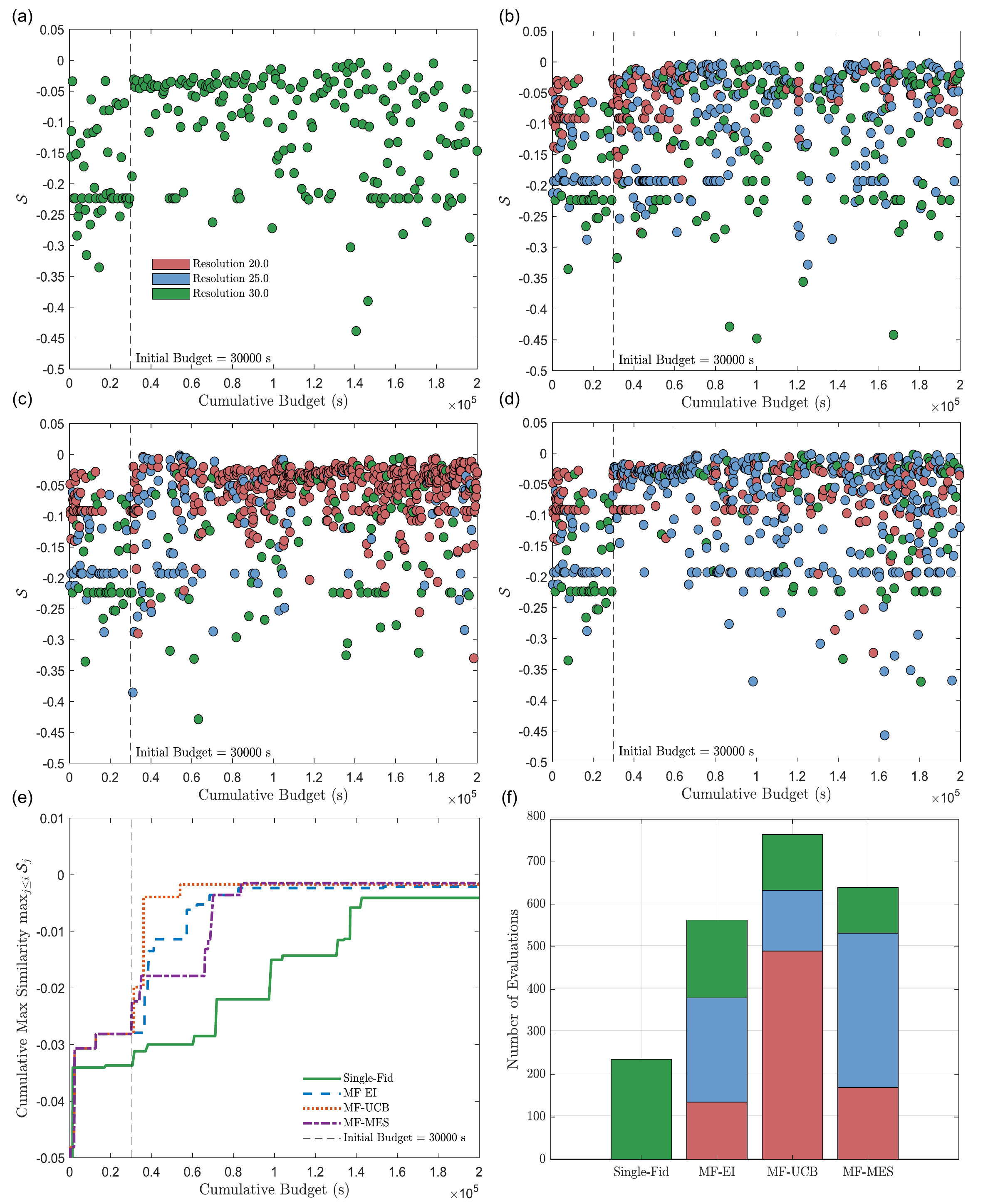}
    \caption[Target 1 -- Evolution of similarity scores and fidelity usage across four optimisation methods.]{{Target 1 -- Evolution of similarity scores and fidelity usage across four optimisation methods. (a-d) Similarity score $\mathcal{S}$ plotted against cumulative computational budget (in seconds) for each method: (a) Single-fidelity, (b) MF-EI, (c) MF-UCB, and (d) MF-MES. Each marker is coloured according to the fidelity level (resolution) at which the corresponding input was evaluated. (e) Comparison of the cumulative maximum similarity score showing their relative optimisation performance under budget constraints. (f) Stacked bar chart indicating the number of evaluations performed at each fidelity level for all methods, illustrating their resource allocation strategies across fidelities.}}
    \label{fig:target_1_optimisation_history}
\end{figure}

{Four methods were employed to reconstruct the desired nonlinear mechanical response, following the workflow described in Sec.~\ref{section:methodology}. For Target 1, the point-wise optimisation history is shown in Fig.~\ref{fig:target_1_optimisation_history}(a-d). In the early iterations, all methods display exploitative behaviour, with low variability in the similarity score following each evaluation. This behaviour is particularly pronounced for the single-fidelity and MF-MES methods, with exploratory behaviour increasing later in the optimisation. This is in line with the greedy nature of the acquisition function used for single-fidelity and MF-MES for determining the input design location, where short-term gains are prioritised by Log-EI. On the other hand, MF-UCB transitions more gradually from exploitation to exploration due to the interaction between its cost-aware acquisition formulation and the limited evaluation budget. With only 35 total evaluations across all fidelities, the surrogate model becomes overconfident early in the optimisation, favouring exploitation through higher-fidelity evaluations. As additional high-fidelity observations are acquired, the inferred input-output landscape is revised, which in turn increases the uncertainty with these additional evaluations. This subsequently shifts the behaviour, favouring exploration through low-fidelity evaluations. It should be noted that since cost is estimated using a GP, additional evaluations also influences its input-output landscape, thus contributing in the shift in behaviour. MF-EI maintains a broader exploration of the design space throughout the process due to its multiplicative, correlation-weighted acquisition function continuing to assign value to uncertain region across the design space, even after promising solutions have been identified. However, such behaviour from MF-EI is not observed for Targets 2 and 3.}

{The convergence trends in Fig.~\ref{fig:target_1_optimisation_history}(e) indicate that all multi-fidelity methods converge more rapidly then the single-fidelity approach when measured against cumulative computational budget. Among them, MF-UCB converges the fastest, followed by MF-EI and MF-MES. This behaviour is likely to be closely linked to the fidelity usage shown in Fig.~\ref{fig:target_1_optimisation_history}(f). MF-UCB strongly favours evaluations at the lowest-cost fidelity level, enabling a larger number of function evaluations within the same budget. MF-EI distributes evaluations more evenly across fidelities, while MF-MES shows a preference for intermediate-to-high fidelity evaluations, particularly the intermediate resolutions.}

\begin{figure}[]
    \centering
    \includegraphics[width=1\textwidth]{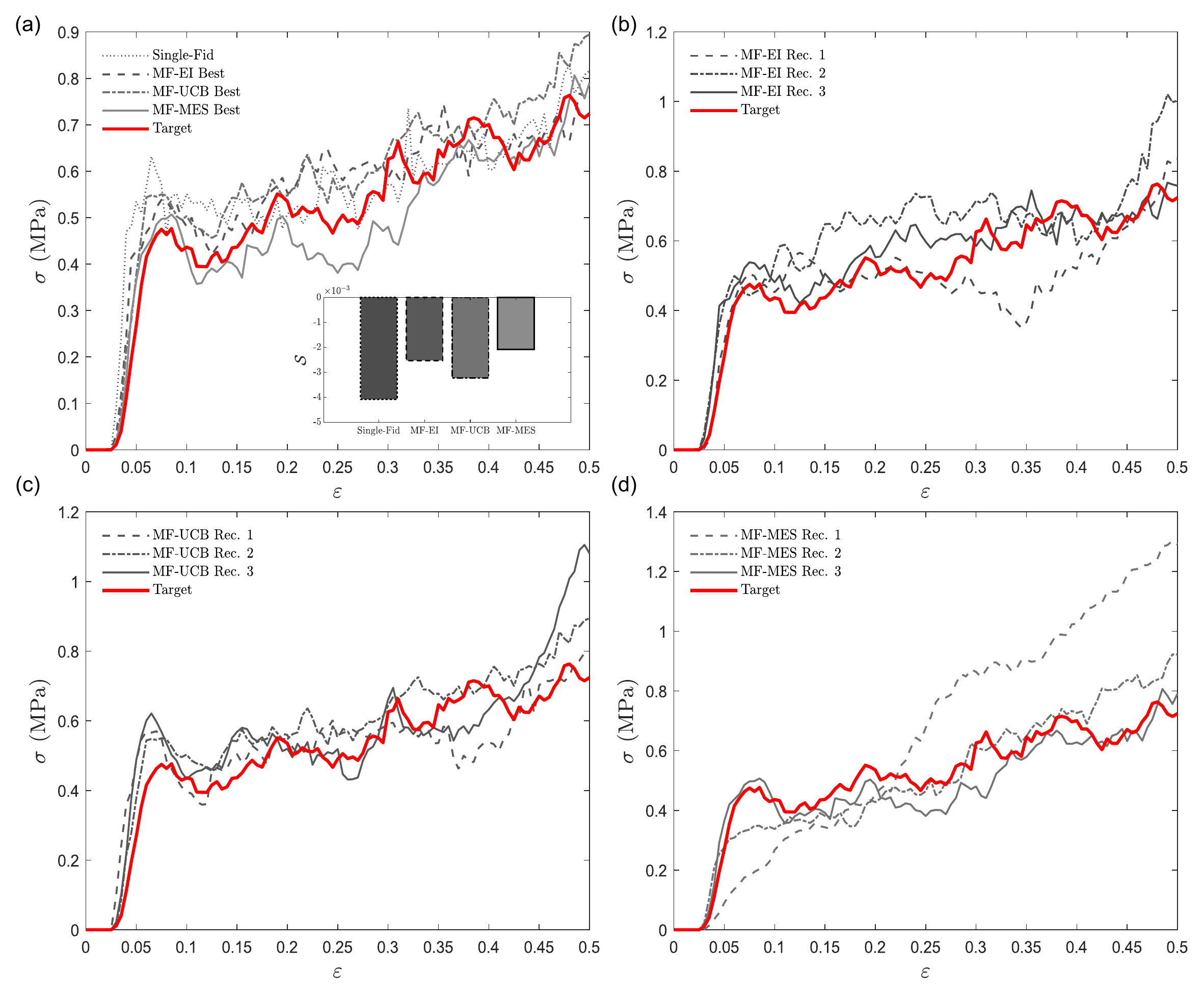}
    \caption{{Target 1 -- (a) Results with the highest similarity score from each method. (b-d) Comparison between the target response and the top three input configurations, based on similarity score, obtained using (b) MF-EI, (c) MF-UCB, and (d) MF-MES. All selected inputs were evaluated at the highest fidelity level.}}
    \label{fig:target_1_combined_results}
\end{figure}

{Two distinct approaches were employed to determine the recommended parameter sets and corresponding similarity scores by minimising the discrepancy between the model output and the target response. In the single-fidelity method, all evaluations were conducted at the highest fidelity level, and the input set yielding the highest similarity score was selected as the optimal solution. In contrast, the multi-fidelity methods operate across three discrete fidelity levels, corresponding to mesh resolutions of 20, 25, and 30, thus generating three unique maximisers, one at each fidelity level.  The resulting optima from each method at each fidelity level are illustrated in Fig.~\ref{fig:target_1_combined_results}. These candidates are subsequently evaluated at the highest fidelity, and the solution with the highest similarity score is selected as the final optimum.}

\begin{table}[h]
\centering
\renewcommand{\arraystretch}{0.6}
\begin{tabular}{@{}cccccc@{}}
\toprule
Method & Rec. \# & $\theta_1$ & $\theta_2$ & $\theta_3$ & $\mathcal{S}$\\
\midrule
\textbf{Target 1} & -- & \textbf{1.69} & \textbf{30.44} & \textbf{6.49} & -- \\

\addlinespace
Single-Fidelity & 1 & 7.54 & 27.2 & 0.8 & \textbf{-0.0041} \\

\addlinespace
\multirow{3}{*}{MF-EI} 
  & 1 & 37.38 & 4.35 & 3.07 &  \textbf{-0.0025}\\
  & 2 & 30.11 & 0 & 10.4 &  -0.0056\\
  & 3 & 27.79 & 12.6 & 6.34 &  -0.0087\\

\addlinespace
\multirow{3}{*}{MF-UCB} 
  & 1 & 39.08 & 0 & 0 &  \textbf{-0.0032}\\
  & 2 & 29.52 & 6.4 & 0.75 &  -0.0041\\
  & 3 & 36.69 & 2.08 & 0.19 &  -0.0059\\

\addlinespace
\multirow{3}{*}{MF-MES} 
  & 1 & 0 & 28.34 & 3.87 & \textbf{-0.0021} \\
  & 2 & 0 & 40.03 & 0 &  -0.004\\
  & 3 & 34.27 & 0 & 12.41 &  -0.041\\
\bottomrule
\end{tabular}
\caption{\label{table:target_1_recommendations} {Target 1 -- Comparison of recommended $\theta$'s obtained from each optimisation method, along with their similarity score $\mathcal{S}$ with respect to the target mechanical response, evaluated at highest fidelity.}}
\end{table}
\label{key}
The design parameter sets yielding the highest similarity scores are summarised in Table.~\ref{table:target_1_recommendations}. The single-fidelity method performs reasonably well, achieving results comparable to the median performance of the multi-fidelity approaches; however, it is consistently outperformed overall, with the multi-fidelity methods providing an average improvement in similarity scores of approximately 37\% for Target 1. Although the optimal parameters is $\boldsymbol{\theta} = (1.69, 30.44, 6.49)$, many recommendations involve configurations where $\theta_1 > \theta_2$ and $\theta_1 \approx 30^{\circ}$. These configurations yield designs that are geometrically identical but oriented perpendicularly within the same plane. This is also a contributing factor in the exploitative behaviour exhibited by the single- and multi-fidelity methods. Full optimisation results can be found in Supplementary material \ref{appendix:Optimisation history MFBO}.

\subsection{{Comparison of single- and multi-fidelity optimisation performance}}

{Building on the illustrative example, the performance of single- and multi-fidelity optimisation strategies is compared across all four target mechanical responses. For each target, the best results obtained by each method, including the corresponding design inputs and the resulting structures, are presented in Fig.~\ref{fig:target_combined_results}. }

\begin{figure}[]
    \centering
    \includegraphics[width=0.8\textwidth]{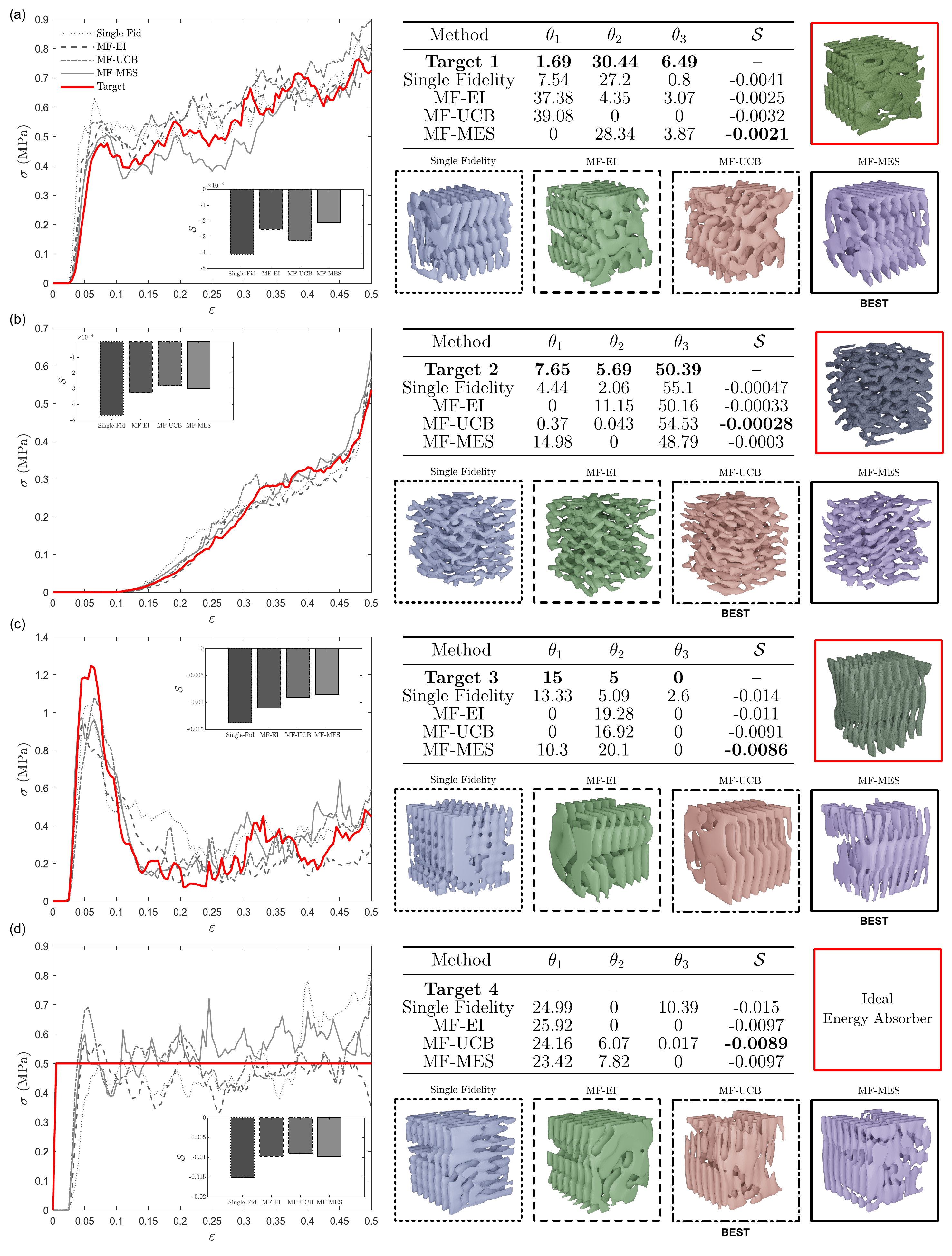}
    \caption{{Summary of inverse design results illustrating the response obtained from four methods across four distinct targets: (a) Target 1, (b) Target 2, (c) Target 3, and (d) Target 4. The inset within the stress–strain graph presents the similarity scores for each method, while the accompanying table lists the design parameters alongside their respective scores. The geometry highlighted in the red square represents the target geometry used to generate the desired response, and the structures shown below the table depict the designs produced with parameters achieving the highest similarity score.}}
    \label{fig:target_combined_results}
\end{figure}

{Across the four targets, all methods achieve low-magnitude similarity scores, indicating that the inverse design framework is generally effective at reproducing the desired mechanical responses. However, as shown by the tables and bar charts in the inset of Fig.~\ref{fig:target_combined_results}, the multi-fidelity methods consistently outperform the single-fidelity benchmark. On average, multi-fidelity optimisation yields an improvement of approximately 35\% in the similarity score relative to the single-fidelity approach. MF-MES and MF-UCB achieve the highest similarity scores for two targets each, followed by the MF-EI and then the single-fidelity method.}

{In terms of convergence speed, multi-fidelity methods demonstrate clear advantages. Similar to the convergence plot, illustrated in Fig.~\ref{fig:target_1_optimisation_history}(e), similar approach was applied to the rest of the targets. Most, if not all, methods for all targets resulted in multi-fidelity approaches converging faster with fewer accumulated cost with higher similarity scores. In addition, the wall time for each inverse design run ranged from approximately 3 to 6 days, depending on the target and the method employed. Single-fidelity and MF-EI runs generally required similar runtimes, while MF-UCB tended to take slightly longer due to its preference for evaluations at the lowest fidelity level, which leads to a larger number of accumulated evaluations. MTGPs scale poorly with increasing dataset size because the covariance matrix, which comprises both task-related and data-related components, is required to be inverted. This slows the subsequent iteration progression. MF-MES required the longest runtime, owing to the computationally expensive acquisition function and its two-step optimisation procedure, which effectively doubles the number of acquisition optimisations required. It should be noted that wall time is distinct from the evaluation budget. Although wall time roughly tracks the evaluation budget, additional time may be needed to tune the model parameters of the complex MTGP kernels or acquisition function. }

{Overall, MF-UCB is recommended when rapid convergence and computational efficiency are priorities, MF-MES is preferable when the highest possible solution accuracy is required and computational resources are available, and MF-EI provides a balanced compromise between these two extremes.}

\subsection{{Mapping target mechanical responses to design parameters}}

{Across all targets, the similarity scores achieved by the optimisation methods vary noticeably, with some cases covering much closer to the desired response than others. This variability is not primarily a consequence of the optimisation strategy itself, b ut instead reflects differences in the underlying input-output relationships between the design parameters $\boldsymbol{\theta}$ and the resulting mechanical response.}

 In particular, when larger values of $\theta$ are required, the resulting structures become more isotropic and thus exhibit similar mechanical responses. This creates a broad region of the design space where many parameter combinations yield comparable responses, making the inverse design process more likely to identify high-similarity solutions if larger $\theta$ values are required to produce the target response. Acquisition functions further reinforce this effect, as their exploratory behaviour often drives evaluations toward the extreme value of $\theta$, increasing the likelihood of stumbling upon valid solutions in these broad regions. In contrast, when smaller $\theta$ values are required, the response is highly sensitive to perturbations, so even minor deviations in $\theta$ lead to very different behaviour and thus lower similarity scores from the optimisation.

This is exemplified using Target 2, where only small values of $\theta_1$ and $\theta_2$ are needed in combination with a larger $\theta_3$ ($> 10$) to produce `Lamellar'-like, more compliant structures. This creates a wide "solution window" in which many parameter combinations yield mechanically similar responses. As a result, the GP can easily capture the trend and effectively traverse the design space, making high-similarity solutions easier to obtain. This contrasts with Targets 1 and 3, where even minor variations in the combination of $\theta$ values result in different solutions.

While similarity scores provide a direct measure of how well the target responses are reproduced, additional insight can be gained by examining the closeness of the recovered input parameters $\boldsymbol{\theta}$ to the original target configurations. For Targets 1-3, where the ground-truth parameter sets are known, this comparison reveals how effectively each method recovers not just the output behaviour but also the underlying design variables. Illustrations of Targets 1 to 3 are shown in Fig.~\ref{fig:target_params}.

When considering only the closeness of the recovered $\boldsymbol{\theta}$ values to the original target configurations (Targets 1-3), all methods perform reasonably well. For Target 1, the recovered parameters were consistently centred around either $\theta_1$ or $\theta_2$, with one of these taking much larger values than the other two. Although the target response was originally defined by $\theta_2 \approx 30^\circ$, some methods instead recovered $\theta_1 \approx 30^\circ$. This arises from the symmetry of the design space, where $\theta_1$ and $\theta_2$ are interchangeable, as they only change the orientation of the geometry along the plane perpendicular to the loading direction and therefore do not alter the structure along the loading direction. Across all methods, the common feature of the recovered sets was that $\theta_3$ remained the smallest parameter, as this directly governs the structure in the loading direction. In terms of parameter closeness, however, MF-MES produced values most consistent with the Target 1 configuration. A similar trend is observed for Target 3, where the desired response corresponds to $\theta_1 = 15^\circ$ and $\theta = 5^\circ$. In contrast, the recovered parameters from the methods often placed greater weight on $\theta_2$ rather than $\theta_1$. Despite this variation, all methods consistently identified that $\theta_3$ must remain very small or zero to reproduce the target response. In terms of parameter closeness, the single-fidelity method yielded values nearest to the target configuration, whereas MF-MES achieved the response most similar to the desired behaviour.

For Target 2, the desired response was achieved with $\theta_3 \approx 50^\circ$ and relatively small values of $\theta_1$ and $\theta_2$. These latter parameters control the stiffness of the layer-like structure: larger $\theta_1$ and $\theta_2$ reinforce adjacent layers with additional material, thereby increasing stiffness in the $x_3$ direction. Among all targets, this case yielded the highest similarity scores across methods, reflecting the broader solution window in which many parameter combinations converge to the desired response. Consistently, all methods recovered values close to $\theta_3 \approx 50^\circ$ with low $\theta_1$ and $\theta_2$. In terms of parameter closeness, MF-EI produced the set most similar to the target, while MF-UCB achieved the highest similarity score. Further discussion on links between design parameters and their effects on mechanical response can be found in Supplementary material \ref{appendix:Linking design parameters to deformation and failure mechanisms}.

\section{Discussion}

Designing cellular composites for realistic working conditions typically demands extensive physical testing or high-fidelity simulations, both of which are time- and resource-intensive. Instead of relying on a trial-and-error or, more generally, forward approaches, inverse design offers a more efficient pathway. This entails directly tailoring the mechanical response of cellular composites for their use in energy-absorbing applications. However, even within an inverse design framework, relying solely on high-fidelity models can be prohibitively expensive. This motivates the incorporation of information from multiple fidelity sources, where evaluations at lower-fidelities act as inexpensive proxies of high-fidelity data, used to guide the search.

In this work, a general framework for the inverse design of spinodoid cellular composites' mechanical response is introduced, which leverages multi-fidelity Bayesian optimisation. Validated finite element simulations that account for both manufacturing defects and fabrication-induced anisotropy were used to construct a fidelity-driven hierarchy. The fidelity levels were defined by mesh resolution, with mesh sensitivity analysis guiding how resolution affects stress-strain behaviour. This hierarchical structure was then modelled using multi-task Gaussian processes, which capture correlations between outputs across fidelity levels. Next, a similarity score was utilised to quantify the difference between the target mechanical response and the best-performing simulated stress-strain response. This effectively reduced the high-dimensional stress vector to a single scalar measure. Crucially, inter-fidelity correlation coefficients were computed to ensure robust knowledge transfer between outputs from different fidelities. It should be noted that the framework is not limited to data from simulations; a mixture of experimental or simulation data can also be utilised. Additionally, the performance of different multi-fidelity strategies, including utility-based and information-theoretic acquisition functions, was benchmarked against a single-fidelity method under an equivalent evaluation budget.

Although this study focused on tailoring the mechanical behaviour of spinodoid cellular composites, the proposed framework is broadly applicable to the inverse design of any scalar property, for instance, energy absorption, or multifunctional properties such as thermal conductivity. Furthermore, the framework extends naturally to function mapping, as demonstrated here, allowing approximations of complete response profiles. Crucially, it can also be applied to investigate the mechanical response under other material behaviours, such as hyperelasticity.

However, it is essential to acknowledge the limitations of the framework. The inverse design methodology provides only approximations of the desired response, and obtaining better representations would require a greater number of evaluations. MTGPs, while effective, scale poorly with data, which may hinder deployment at larger scales unless more efficient surrogates are employed. Sparse GPs, for instance, reduce computational complexity by introducing inducing points \cite{titsias2009variational, liu2020gaussian}. In addition, the choice of similarity function, or the weighting of specific strain ranges, can bias the inverse design process towards different features of the response, leading to different optimal designs. Similarly, the choice of GP kernel for modelling the design parameters shapes the representation of the design space and, in turn, the trajectory of the inverse design process. In principle, the influence of kernel hyperparameters and initialisation heuristics could be mitigated by repeating the inverse design process with different Sobol' sequence seeds for the initial data. However, this was not pursued here, as the primary objective was to highlight the plug-and-play nature of the proposed framework while keeping the evaluation cost minimal. Nonetheless, the study relies on validated FEM simulations, which, while robust, may not capture all sources of uncertainty. Experimental results typically show variation in stress-strain response due to fabrication imperfections or testing variability. As a result, utilising this framework with both experimental data and multi-fidelity simulation data would provide a good test for real-world deployment.

Another limitation arises from the target-specific nature of the inverse design process. This stems from the need to transform the objective function using a similarity score relative to a given target. If a surrogate model is trained directly on this transformed objective, then each new target response requires re-optimising the surrogate model parameters, which is a computationally expensive task, particularly for MTGPs, as discussed earlier. An alternative is to decouple the surrogate from the target. This involves adopting dimensionality reduction techniques such as PCA or autoencoders to obtain a latent representation of the stress-strain response. Surrogate models are fit on this latent representation rather than the full response. The target response is also projected into the same latent space. By sampling the posterior of the surrogate model, the difference between the target and sampled surrogate posterior values can then be found. These transformed posterior samples are then passed to the acquisition function to identify the next sampling point. While the drawbacks of this approach have been discussed, it could provide valuable opportunities for future studies.

For high-dimensional designs, such as graded spinodoids (further details on the method for generating graded spinodoids in \ref{appendix:Extension to Graded Spinodoids}), this expansion significantly increases the input space. Importantly, graded spinodoids have been shown to exhibit greater energy absorption than their homogeneous counterparts \cite{liu2024mechanical}, highlighting the practical benefit of exploring such complex architectures. {While the present study reduced the design space, retaining the full design space or introducing functional grading would substantially increase the dimensionality of the optimisation problem, which is known to slow convergence and increase sensitivity to the initial dataset. Insights from the synthetic multi-fidelity benchmarks already highlight this behaviour: higher-dimensional functions exhibit greater run-to-run variability and slower convergence unless the initial dataset provides sufficiently rich coverage of the design space. Nevertheless, these benchmarks also demonstrate that multi-fidelity strategies remain advantageous in higher-dimensional settings, consistently outperforming single-fidelity optimisation by leveraging inexpensive low-fidelity evaluations to guide exploration, especially in the case of MF-EI. When extending the framework to fully parameterised or graded spinodoid architectures, additional scalability measures would therefore be required.} Possible remedies include dimensionality reduction via Sobol' sensitivity analysis \cite{saltelli2010variance, sobol2001global, Kansara:MOBO(2025), guo2026multi}, by exploiting known optima around previously evaluated points to form trust regions (TuRBO) \cite{eriksson2019scalable}, or utilising random embedding strategies to allow high-dimensional problems to be represented in lower-dimensional subspaces \cite{Wang2013}.

{Future work could also extend the finite element models by incorporating strain-rate dependency through models such as Johnson-Cook plasticity, or a viscoelastic model with fracture behaviour, allowing for more complex material behaviour to be captured. Additionally, while the present study addresses a single scale or mapped function, many applications require inverse designing multiple objectives simultaneously. This could be achieved by replacing the Log-EI-based two-step approach in the MF-MES method with a multi-objective acquisition function, such as Expected Hypervolume Improvement (EHVI) \cite{daulton2020differentiable}, multi-objective knowledge gradient \cite{daulton2023hypervolume}, or entropy-based methods \cite{belakaria2020multi} to extend the framework to accommodate multi-objective multi-fidelity. Finally, although demonstrated here at the unit-cell level, future studies could implement the proposed strategy on a macroscale or within multi-scale structures to assess its efficacy.}

\section{{Conclusion}}

{This work introduces a general and flexible framework for the inverse design of spinodoid cellular composites using multi-fidelity Bayesian optimisation. By combining validated finite element simulations, a fidelity-driven hierarchy was adopted in tandem with MTGPs capable of modelling inter-fidelity correlations. The proposed framework provides an efficient and robust approach for tailoring complex mechanical responses. A similarity-based objective function enabled the comparison between simulated and target stress-strain curves, allowing the inverse design problem to be reformulated as a scalar optimisation task without loss of information.}

{Four representative target responses were selected, along with four optimisation strategies. The results showed that multi-fidelity approaches consistently outperformed the single-fidelity method, achieving, on average, 35\% higher similarity scores. This demonstrates that leveraging low-fidelity evaluations to guide high-fidelity exploration yields more accurate and efficient recovery of target behaviours. Moreover, the framework successfully identified not only stress-strain responses that closely match the targets but also input parameter combinations that reflect the underlying deformation mechanisms. Despite its effectiveness, the framework has limitations. MTGPs scale poorly with dataset size, and optimisation outcomes remain sensitive to the similarity metric, kernel choice, and initial dataset. Target-specific surrogate training also limits reusability across different inverse design tasks. Finally, extending the framework to graded or heterogeneous spinodoids will require strategies to mitigate high dimensionality.}

{Overall, the proposed MFBO framework establishes a powerful and generalisable foundation for the inverse design of architected materials. While challenges related to scalability, surrogate modelling, and target-specific optimisation remain, the results demonstrate the potential of multi-fidelity strategies to accelerate the discovery of high-performance cellular composites and to expand the accessible design space.}

\section{Acknowledgement}
W. Tan acknowledges the financial support from the EPSRC New Investigator Award (grant No. EP/V049259/1). This work was selected by the ERC and funded by UK Research and Innovation (UKRI) under the UK government’s Horizon Europe funding guarantee (No. EP$\slash$Y037103$\slash$1). The authors extend much gratitude to Dr. Miguel A. Bessa and Dr. Siamak F. Khosroshahi for the valuable discussions, insights, and support throughout the project leading to this manuscript.

\section{Data availability}
The data and codes needed to reproduce and evaluate the work of this paper are available in the GitHub repository, once this manuscript is published: https://github.com/MCM-QMUL.

\bibliographystyle{elsarticle-num}
\bibliography{Ref_anon}

\setcounter{figure}{0}
\renewcommand{\thefigure}{S\arabic{figure}}
\setcounter{section}{0}
\renewcommand{\thesection}{S\arabic{section}}
\setcounter{table}{0}
\renewcommand{\thetable}{S\arabic{table}}
\captionsetup[figure]{labelformat=simple, labelsep=colon}

\section*{Supplementary Information}

\section{FEM setup and experimental validation}
\label{appendix:FEM setup and experimental validation}
FEM simulations were employed to investigate the mechanical response of the generated spinodoid structures. Structures with dimensions of 40 mm $\times$ 40 mm $\times$ 40 mm were created using a workflow combining MATLAB for geometry generation and ABAQUS for simulation. These analyses were performed using ABAQUS Explicit, with the structures meshed using C3D4-type tetrahedral elements. To study the behaviour of spinodoid cellular composites under compression, two rigid plates were introduced into the FEM model, one positioned at the top and the other at the bottom of the spinodoid. The bottom plate acted as a fixed anvil, while the top plate served as the loading surface, subjected to a displacement-controlled boundary condition. The structure was compressed in the $x_3$ direction until a strain of 50\% was reached, as indicated by the datum axes in Fig.~\ref{fig:spinodoids}. A general contact interaction was defined between the plates and the structure, allowing densification while preventing self-contact during deformation. Owing to the symmetry of the design parameters $\boldsymbol{\theta}$, uniaxial compression in a single direction was sufficient to characterise the structural response across the entire design space.

{Since the structures were fabricated using 3D printing, a constitutive model was developed to accurately represent the anisotropic mechanical behaviour. This model was calibrated using material characterisation data.} Cubic samples (20 mm $\times$ 20 mm $\times$ 20 mm) were tested in compression, and ISO-527 type A dogbone specimens were tested in tension. Each sample type was printed in three orientations, $0^\circ$, $45^\circ$, and $90^\circ$, relative to the loading direction, to quantify the anisotropy introduced by the material extrusion process. The resulting stress-strain curves from these tests are presented in Fig.~\ref{fig:carbon_p_experimental_validation}(a). All test specimens were printed using short carbon fibre-reinforced PET-G (Carbon-P) filament, containing 20\% fibre by weight as specified by the manufacturer. As mentioned, the printing process inherently introduces directional mechanical behaviour due to fibre alignment and inter-layer bonding. To represent this anisotropy in simulation, the material was modelled as orthotropic, with elastic properties defined directionally using experimentally determined engineering constants. Anisotropic plasticity was implemented through the quadratic Hill yield criterion, and a ductile damage initiation model was included to simulate the onset of material failure, with damage evolution to describe the stiffness degradation through fracture energy. The FEM setup adopted was in line with the study \cite{Kansara:MOBO(2025)}, with the key difference being the addition of damage criteria, damage evolution, and the use of short carbon fibre reinforced PET-G to fabricate the cellular composite. 

\begin{figure}[!h]
    \centering
    \includegraphics[width=1\textwidth]{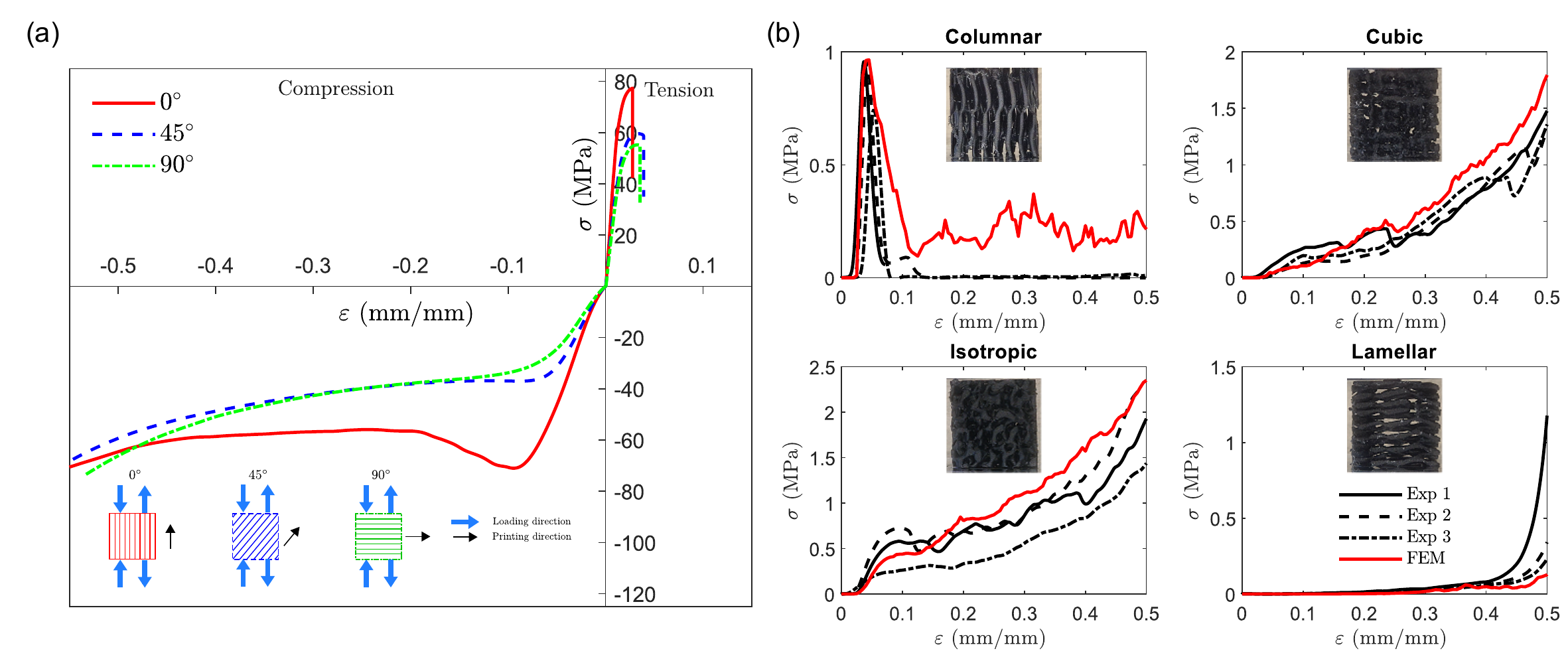}
    \caption{Results from conducting: (a) Tensile and compression tests on cubes and dogbone samples printed using Carbon-P; (b) Validation of FEM results using 3D printed spinodoids with geometry described in Fig.~\ref{fig:spinodoids}.}
    \label{fig:carbon_p_experimental_validation}
\end{figure}

A mesh convergence study was first performed to ensure mesh-independent results, demonstrating convergence at a resolution of 30 and higher. Following this, {an initial validation of the constitutive model was performed by simulating the compression of a cube printed at 90$^{\circ}$ \textcolor{black}{(found in Supplementary Information ~\ref{appendix:Material characterisation validation and Mesh convergence})}}. Selected spinodoid topologies (as shown in Fig.~\ref{fig:spinodoids}) were used to further validate the FEM model. Unlike the smaller cubic samples, spinodoid prototypes measuring 40 mm $\times$ 40 mm $\times$ 40 mm were fabricated using an UltiMaker S5 with Carbon-P filament and water-soluble PVA support material. These structures were subjected to compressive testing under quasi-static loading at a strain rate of 0.1 $\mathrm{s}^{-1}$, consistent with the conditions used in material characterisation. The results from experiments and simulations have been plotted in Fig.~\ref{fig:carbon_p_experimental_validation}(b). Furthermore, the simulation of spinodoid structures used plasticity data derived from the 90$^\circ$ cube tests and elastic constants from the compression experiments. However, this initially resulted in overly stiff structural responses, leading to discrepancies between the experimental and simulated results. To address this, the material stiffness parameters were reduced to 30\% of their original values. This adjustment accounts for imperfections introduced during fabrication, such as poor inter-layer bonding and structural defects caused by overhanging features, which can reduce the effective mechanical performance of printed parts. After applying this correction, the simulations showed good agreement with experimental data. \textcolor{black}{Further detail can be found in Supplementary Information \ref{appendix:Effect of stiffness on stress-strain response}}. For the `Columnar' geometry, the brittle nature of the failure mode caused complete loss of platen contact after the initial break in all repetitions, resulting in zero stress values post-failure and a noticeable disparity between simulation and experiment. To reduce computational cost, a combination of time and mass scaling was applied. \textcolor{black}{The influence of these parameters on the initial stress response is detailed in Supplementary Information \ref{appenedix:Effect of time and mass scaling on mechanical response}}. This adjustment explains the initial zero-stress period observed in the simulations.

Despite these adjustments, some discrepancies between experimental and simulated post-yield behaviour remained. This limitation stems from the fracture energy parameter ($G_c$) used in the damage evolution model, which governs stiffness degradation and energy dissipation during failure. In this study, $G_c$ was approximated from the manufacturer's specified Charpy notched impact energy. While the Charpy test provides an indication of impact toughness, it is not a standardised measure of fracture toughness and does not directly translate to fracture energy under quasi-static loading conditions. Consequently, the use of this approximation introduces uncertainty in modelling the damage process. {Following stiffness penalisation, the effect of fracture energy, $G_c$, on the predicted response was examined through a sensitivity study spanning $0.1-1 \times G_c$ for two representative spinodoid topologies. As $G_c$ controls the onset of material damage, this analysis evaluates its role in shaping the post-yield behaviour. \textcolor{black}{Detailed results are presented in Supplementary Information \ref{appendix:Effect of fracture energy on stress-strain response}.}}

\section{Material characterisation validation and mesh convergence}
\label{appendix:Material characterisation validation and Mesh convergence}

Fig.~\ref{fig:cube_validation_and_effect_of_resolution_on_stress_strain}(a) presents a comparison between the results of FEM simulations and material characterisation tests. The stress-strain curves demonstrate good overall agreement, with the FEM model accurately capturing both the initial elastic response and the densification behaviour at higher strains. However, a slight discrepancy is observed at the onset of plastic deformation. This deviation may be attributed to the assumptions involved in determining the coefficients for the Hill anisotropic yield criterion.

\begin{figure}[ht]
    \centering
    \includegraphics[width=1\textwidth]{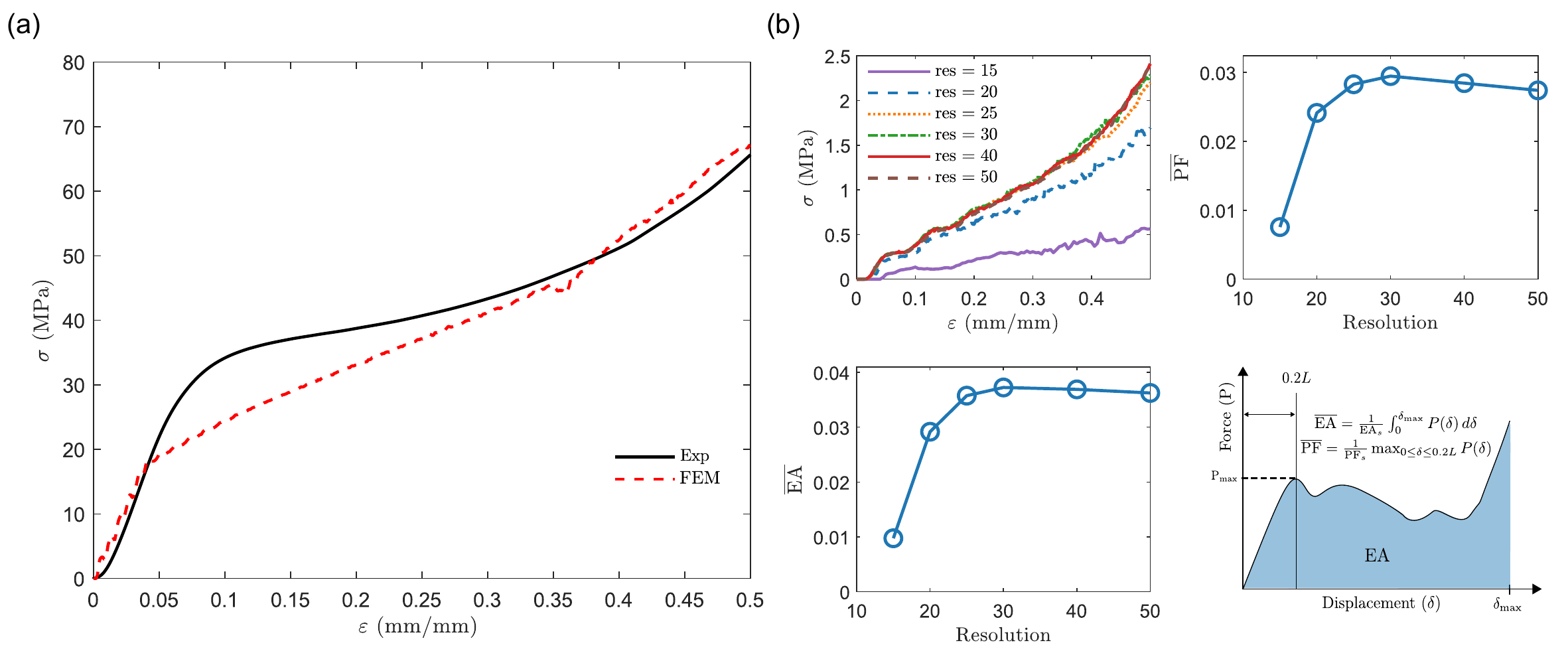}
    \caption[Comparison between experiment and FEM simulation for compression of a 90$^\circ$ cube used for material characterisation. Additional plots showing the effect of resolution on the stress-strain behaviour, and normalised EA and PF.]{(a) Comparison between experiment and FEM simulation for compression of a 90$^\circ$ cube used for material characterisation. (b) (Clockwise) Stress-strain behaviour of `Isotropic' spinodoid generated using various mesh resolutions; Graph plotting the effect of resolution on the normalised peak force; Graph illustrating the method used to calculate the peak force and energy absorption where EA$_s$ and PF$_s$ indicates the EA and PF values for an equivalently sized cube made from the equivalent material; The effect of resolution on the normalised energy absorption are plotted.}
    \label{fig:cube_validation_and_effect_of_resolution_on_stress_strain}
\end{figure}

The top-left subplot in Fig.~\ref{fig:cube_validation_and_effect_of_resolution_on_stress_strain}(b) demonstrates how resolution affects the stress-strain response of the `Isotropic' spinodoid. It is clear that resolution has a significant impact because it directly determines the level of geometric detail generated by the pipeline. When large elements are used, a low resolution fails to accurately capture the prescribed relative density. Consequently, the mechanical response of the structure under compression becomes softer. This is further reinforced by the normalised $\overline{\mathrm{PF}}$ and $\overline{\mathrm{EA}}$ plotted against the resolution, shown in the top-right and bottom-left subplots, respectively. Together, these plots serve as a mesh convergence study to guide the selection of appropriate fidelity levels for the subsequent multi-fidelity Bayesian optimisation process.

Both $\overline{\mathrm{PF}}$ and $\overline{\mathrm{EA}}$ indicate that the performance metrics begin to stabilise at a resolution of 30. In this analysis, $\overline{\mathrm{EA}}$ was chosen as the primary indicator of convergence, as it is derived from integrating the area under the force-displacement curve, offering a more robust representation of the stress-strain behaviour. In contrast, $\overline{\mathrm{PF}}$ was not used as a convergence criterion, since the normalised peak force continues to decrease beyond a resolution of 30, which is likely due to a combination of improved numerical stability and reduced discretisation error, albeit at the cost of significantly increased simulation time.

Ultimately, a resolution of 30 was selected as the high-fidelity ground truth, balancing accuracy and computational cost. For the lower-fidelity levels, resolutions of 20 and 25 were chosen, as they provide a reasonable approximation of the high-fidelity results.

\section{Effect of stiffness on stress-strain response}
\label{appendix:Effect of stiffness on stress-strain response}

To align the initial elastic response and peak stress from FEM simulations with experimental observations, a reduction factor was applied to the elastic properties obtained from material characterisation tests. This adjustment targeted the orthotropic material constants, based on the assessment of the stiffness reduction required. A percentage-based penalty was introduced and explored over a range from $0.1E$ to $E$, where $E$ denotes the elastic modulus corresponding to the structure's printing orientation. The full results are plotted in Fig.~\ref{fig:effect_of_stiffness}. The stiffness reduction accounted for all fabrication defects, resulting in poor inter-layer bonding and adhesion. This allows layers to slide under compression, reducing the effective shear stress transfer between layers. In addition, poor calibration of the print bed can also lead to cavitation in between layers, leading to a reduced load-bearing capacity, causing local stress concentrations, which lowers the stiffness and strength. From the figure, it is clear that scaling the elastic properties down to 30\% of their initial values sufficiently captures the experimental behaviour.

\begin{figure}[ht]
    \centering
    \includegraphics[width=1\textwidth]{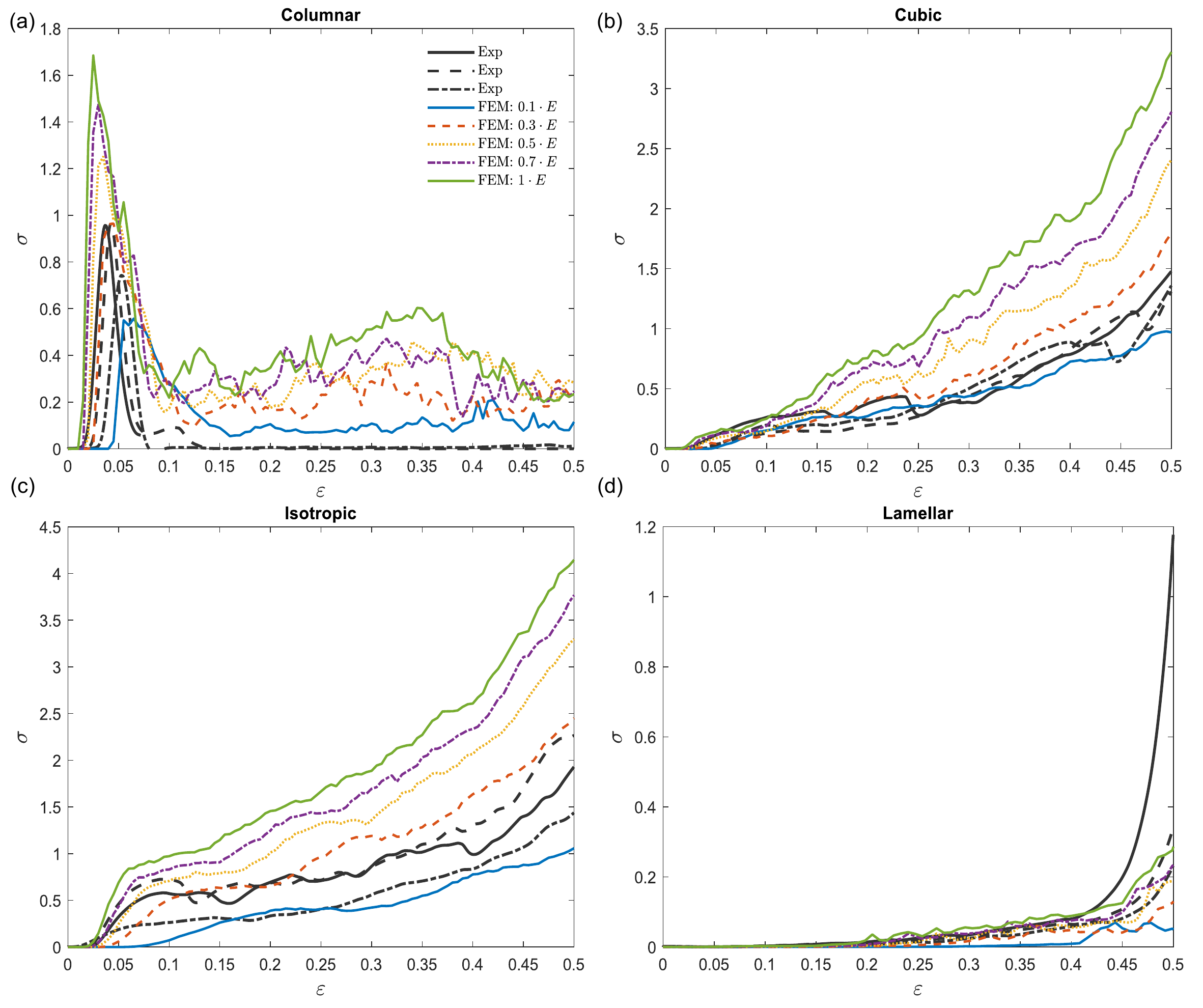}
    \caption[Effect of stiffness on the mechanical response.]{Plots illustrating the effect of varying stiffness levels on four test topologies, with $E$ representing the elastic modulus, alongside a comparison with experimental results.}
\label{fig:effect_of_stiffness}
\end{figure}

\section{Effect of time and mass scaling on mechanical response}
\label{appenedix:Effect of time and mass scaling on mechanical response}
To further reduce the simulation time, two approaches are available in ABAQUS, both of which influence the size of the stable time increment and are therefore particularly relevant when modelling quasi-static loading. In this study, a total time interval of 0.003 s was specified, and a mass scaling with a factor of two was applied. The effect of these choices on the stress-strain curves is plotted in Fig.~\ref{fig:mass_and_time_scaling}, where `Isotropic' structure was used as the test case.

Here, time scaling reduces the number of increments required to reach a given total simulation time, while mass scaling increases the stable time increment by artificially raising the material density and thereby lowering the wave propagation speed. Although this reduces computational cost, it also delays the arrival of stress waves at the anvil, where the reaction force is extracted, meaning the reaction force remains negligible until the wave has traversed the structure. This is the reason for the initial zero stress at the beginning.

\begin{figure}[ht]
    \centering
    \includegraphics[width=1\textwidth]{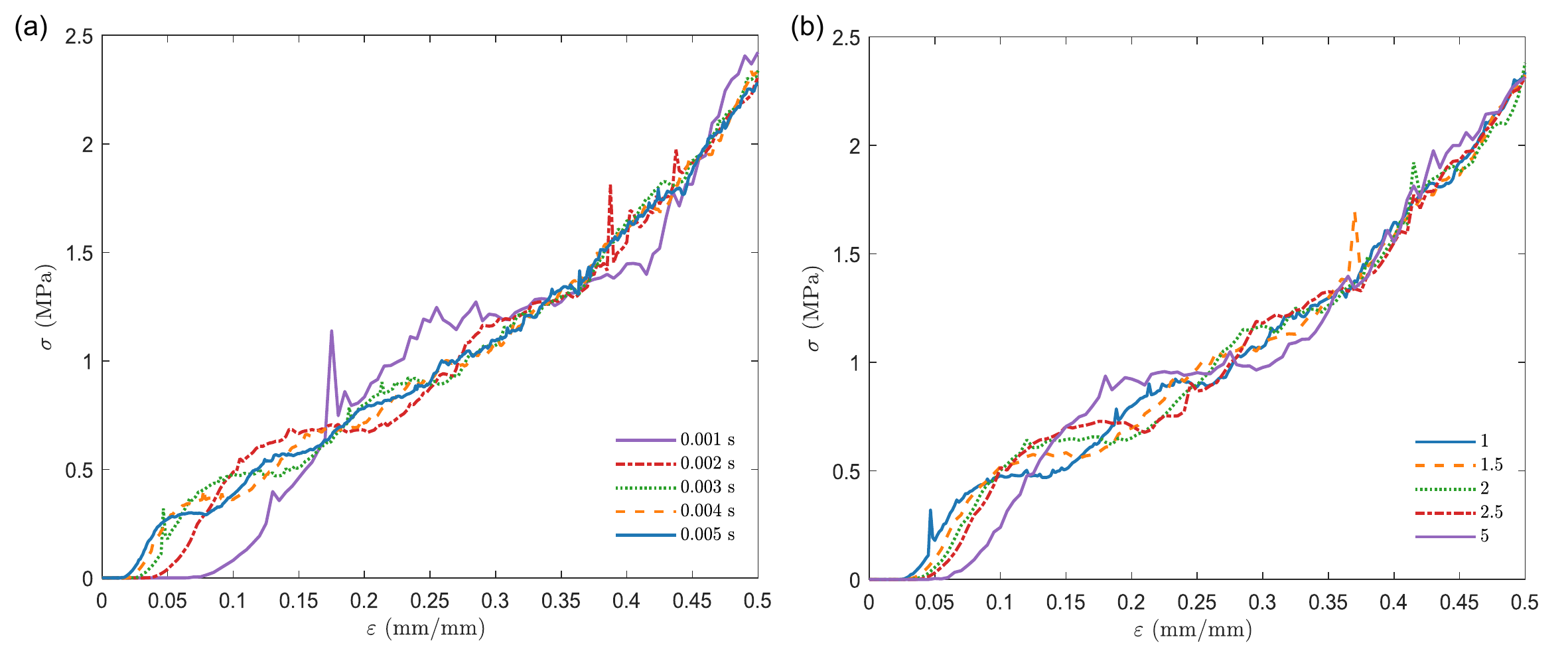}
    \caption[Effect of time and mass scaling on the mechanical response.]{Plots showing the effect of (a) time scaling and (b) mass scaling on `Isotropic' spinodoid geometry, generated with resolution 30, on the stress-strain curves.}
\label{fig:mass_and_time_scaling}
\end{figure}

\section{Effect of fracture energy on stress-strain response}
\label{appendix:Effect of fracture energy on stress-strain response}

The fracture energy parameter plays a role in the mechanical response of the structures. Because its value was taken from the manufacturer's filament specification, it is necessary to examine its influence on structural behaviour. To this end, simulations were conducted at three fracture energy levels: $0.1 \times$, $1 \times$, and $10 \times$ the manufacturer-provided value. As shown in Fig.~\ref{fig:effect_of_damage}, the impact of fracture energy varies across topologies, particularly in the post-yield region. For the `Columnar' structure (Fig.~\ref{fig:effect_of_damage}(a)), the initial response is largely insensitive to fracture energy, as the initial load drop is geometry-driven, as it may be better at redistributing the stress, likely associated with buckling and elastic-plastic instability. Beyond the peak stress, however, differences emerge in post-yield behaviour, with higher fracture energies resulting in greater energy absorption. This trend is illustrated quantitatively in the inset plot, where energy absorption increases with fracture energy, although the variation remains modest compared to the response observed for the `Isotropic' topology. 

Fig.~\ref{fig:effect_of_damage}(b) demonstrates that the `Isotropic' structure is highly sensitive to fracture energy. This sensitivity arises because the topology promotes stress localisation due to interconnectivity, leading to premature failure at interconnecting regions. The inset shows a substantial variation in energy absorption across the three fracture energy levels, highlighting that accurate prediction of post-yield behaviour is strongly dependent on the correct value of $G_c$. Consequently, precise calibration of fracture energy is essential for robust simulations of damage-driven responses.

\begin{figure}[ht]
    \centering
    \includegraphics[width=1\textwidth]{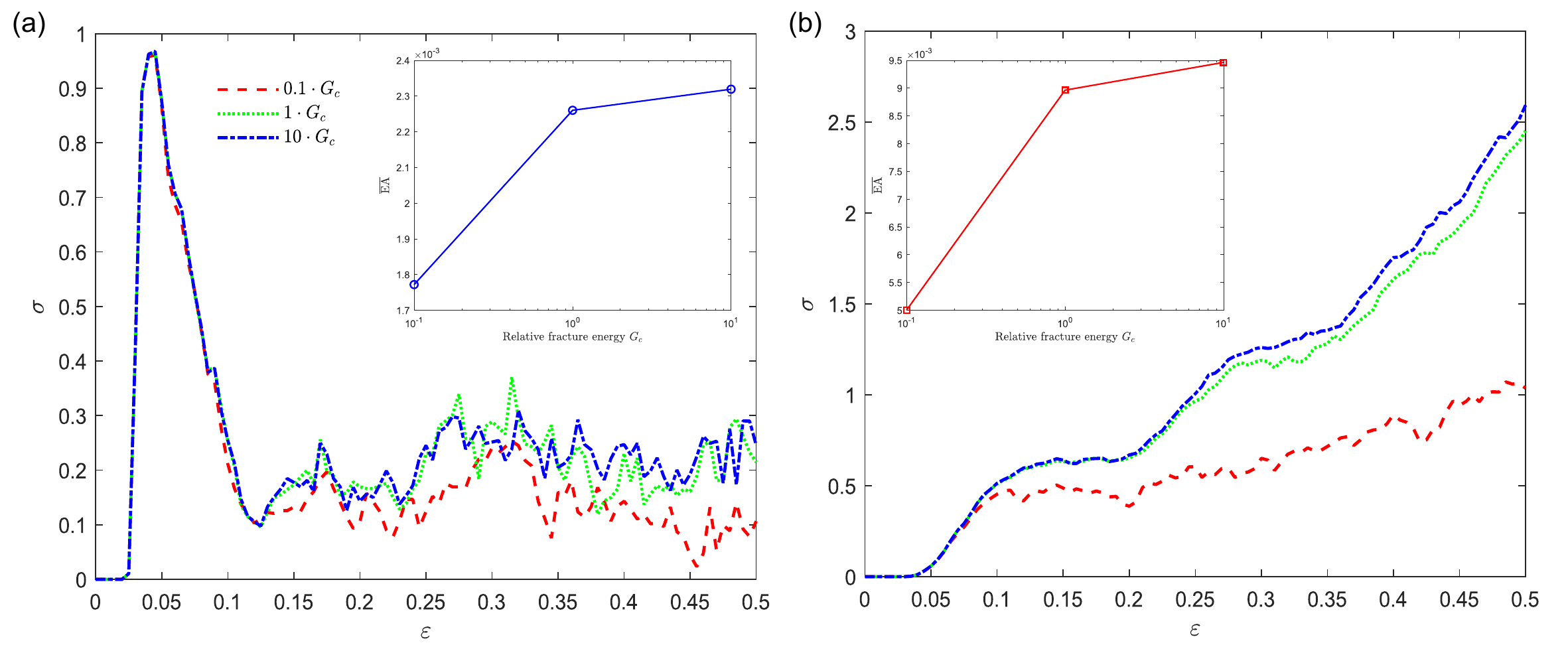}
    \caption[Effect of fracture energy on the mechanical response.]{Plots showing the effect of various fracture energies on the mechanical response of (a) `Columnar' and (b) `Isotropic' spinodoid geometries. The insets show the effect of various $G_c$ on the normalised energy absorption.}
\label{fig:effect_of_damage}
\end{figure}

\section{Pearson correlation coefficients for targets 1, 2, 3, and 4}
\label{appendix:Pearson correlation coefficients for Targets 2, 3, and 4}

\textcolor{black}{To further validate the suitability of using multiple fidelity levels within a single optimisation framework, correlation coefficients were computed between simulation outputs at different mesh resolutions. An initial set of 35 design points was generated using a Sobol' sequence and evaluated at all fidelity levels. The distribution across the input space can be seen in the pair plot illustration in Fig.~\ref{fig:initial_data_sim_time_correlation}(a). Here, only a small amount of data was generated to minimise the computation cost whilst also allowing a sufficiently accurate comparison between fidelity outputs to be made.}

\textcolor{black}{For MTGP to be successful, it is essential for there to be a transfer of information. One way to assess the potential for effective inter-fidelity knowledge transfer is by computing the Pearson correlation coefficient, which plays a central role in the MTGP framework. High correlation between a lower-fidelity model and its high-fidelity counterpart suggests that meaningful knowledge transfer is feasible. To quantify this, the initial datasets across fidelity levels were compared against the high-fidelity outputs, as shown in Fig.~\ref{fig:initial_data_sim_time_correlation}(b) and (c). For resolution 20 vs.~30, the Pearson correlation coefficient was approximately 0.89, indicating a strong positive correlation. However, the distribution of points shows a consistent deviation below the $x=y$ line, suggesting a systemic bias likely caused by the lower resolution underestimating the mechanical response. This bias introduces additional structured noise, which varies with different fidelity, that the MTGP is designed to model. In contrast, resolution 25 vs.~30 yielded a higher Pearson correlation coefficient of $\sim 0.98$, with points more tightly clustered around the identity line, although a slight underestimation remains, again likely due to the generation of incomplete structure features. These particular values of $\tilde{\rho}$ are for similarity scores calculated with respect to Target 1; correlations for the rest of the targets can be found in Fig.~\ref{fig:correlations_target_2_to_4}. Unlike multi-fidelity methods, which often rely on complex models like MTGP, the single-fidelity optimisation approach used here employs a standard GP as the surrogate model. Although an MTGP could also be applied in the single-fidelity setting, it would result in additional computational overhead from model fitting, on top of the evaluation cost.}

\begin{figure}[!h]
    \centering
    \includegraphics[width=1\textwidth]{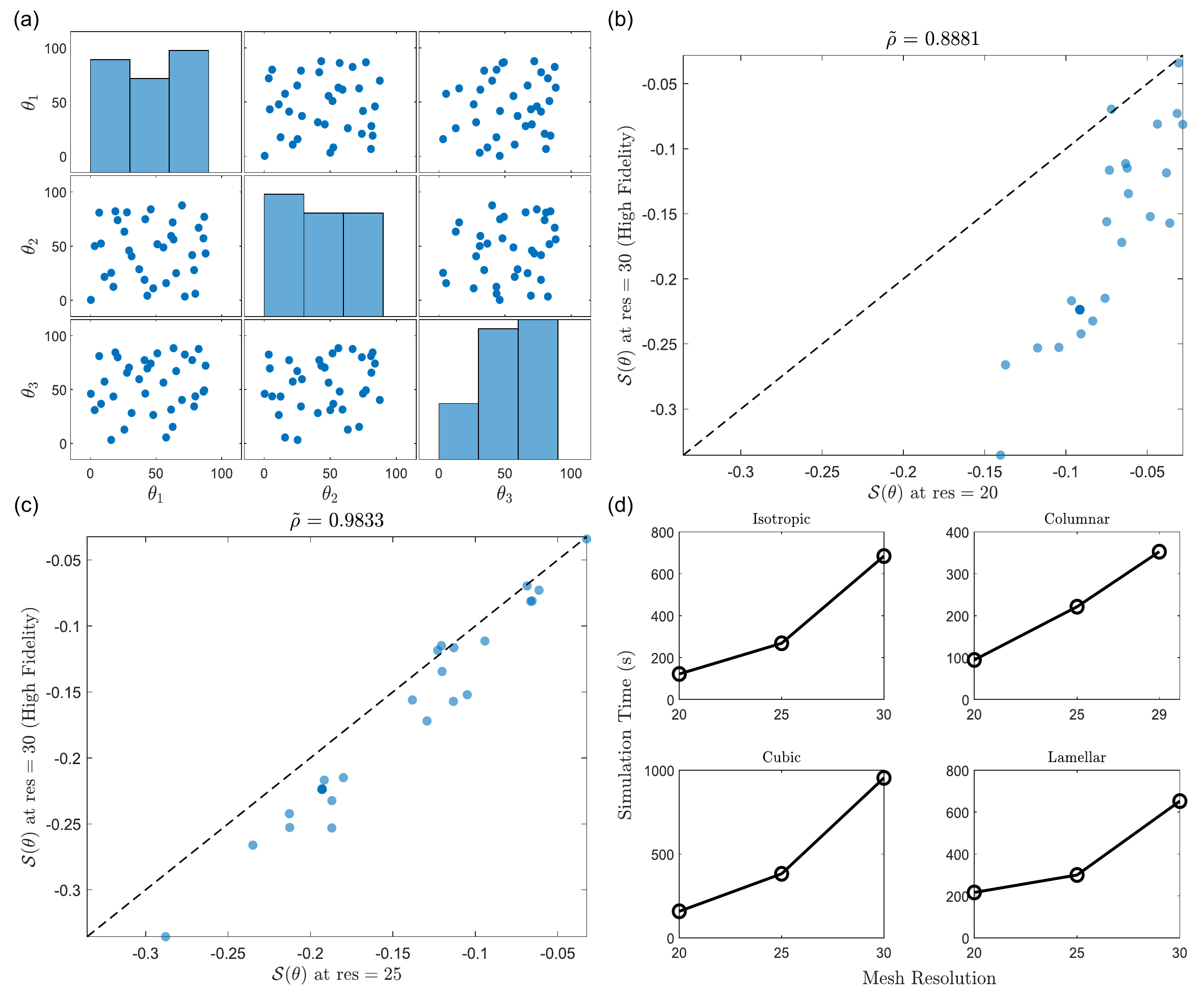}
    \caption{\textcolor{black}{(a) Pair plot of the initial design dataset consisting of 35 points sampled across the input space. Each design (i.e., set of $\boldsymbol{\theta}$ values) was evaluated at three different fidelity levels.  (b) and (c) show the Pearson correlation plots comprising similarity scores obtained at mesh resolutions of 20 vs.~30 and 25 vs.~30, respectively, for Target 1, illustrating agreement across fidelity levels.  (d) Visualisation of how mesh resolution impacts runtime across four test cases shown in Fig.\ref{fig:spinodoids}.}}
    \label{fig:initial_data_sim_time_correlation}
\end{figure}

\textcolor{black}{Fig.~\ref{fig:correlations_target_2_to_4} shows the Pearson correlation coefficients obtained using the input dataset visualised in Fig.~\ref{fig:initial_data_sim_time_correlation}(a), with similarity scores calculated with respect to the targets plotted in Fig.~\ref{fig:target_params}.}

\begin{figure}[]
    \centering
    \includegraphics[width=1\textwidth]{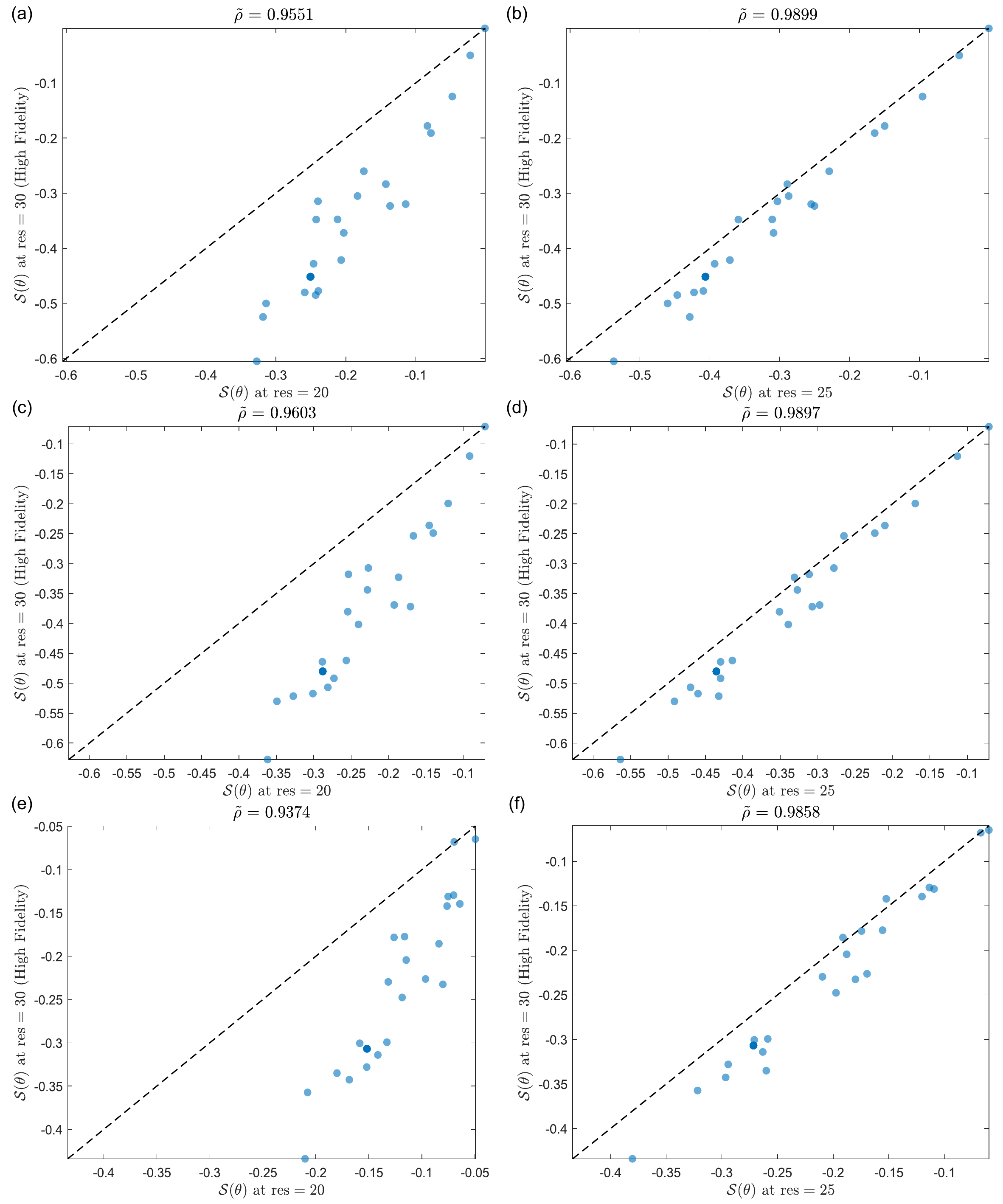}
    \caption{Each row presents the Pearson correlation coefficients comparing mesh resolutions 20 vs.~30 and 25 vs.~30 for different target responses, respectively. Subfigures (a) and (b) correspond to Target 2, (c) and (d) to Target 3, and (e) and (f) to Target 4.}
    \label{fig:correlations_target_2_to_4}
\end{figure}

\section{\textcolor{black}{Further MTGP details and other multi-output GPs}}
\label{appendix: Further MTGP details and other multi-output GPs}

\textcolor{black}{Let the objective function at fidelity level  $s$  be denoted as  $f^{s}(\mathbf{x})$, where $\mathbf{x} \in \mathbb{R}^q$ is the input vector in a $q$-dimensional design space, and $s \in \{1,...,S\}$ indexes the fidelity levels (with $s = S$ being the highest fidelity). The resulting joint prior over all fidelity levels is therefore given by}
\begin{equation}
    \textcolor{black}{\mathbf{f}(\textbf{x}) \sim \mathcal{GP}(\textbf{m}(\textbf{x}),\mathcal{K}(\textbf{x},\textbf{x}'))}
\end{equation}
\textcolor{black}{where $\mathbf{f}(\textbf{x}) = [f^{(1)}(\textbf{x}),...,f^{(S)}(\textbf{x})]^\top$, and $\textbf{m}(\textbf{x}) = [m^{(1)}(\textbf{x}),...,m^{(S)}(\textbf{x})]^\top$, typically assumed to be zero, with outputs being normalised and standardised. }

\textcolor{black}{In the scalar-output GP, the kernel matrix is given by}
\begin{equation}
    \textcolor{black}{\mathbf{K} = [k(\mathbf{x}_i, \mathbf{x}_j)]_{i,j=1}^N \in \mathbb{R}^{N \times N}}
\end{equation}
\textcolor{black}{where $N$ is the total number of training points.}

\textcolor{black}{In contrast, for a multi-output GP (MOGP) with $S$ outputs (e.g., fidelity levels), the kernel becomes matrix-valued. Suppose fidelity level $s$ has $N_s$ training inputs}
\begin{equation}
    \textcolor{black}{\textbf{X}_s = \{\textbf{x}^1_s,...,\textbf{x}^{N_s}_s\}, \quad s\in\{1,2,...,S\}}
\end{equation}
\textcolor{black}{Then the full covariance matrix over all outputs and training inputs is block-structured}

\begin{equation}
\textcolor{black}{\mathbf{K}_\text{MOGP} =
\begin{bmatrix}
\mathbf{K}_{1,1}(\mathbf{X}_{1},\mathbf{X}_{1}) & 
\mathbf{K}_{1,2}(\mathbf{X}_{1},\mathbf{X}_{2}) & 
\cdots & 
\mathbf{K}_{1,S}(\mathbf{X}_{1},\mathbf{X}_{S}) \\[6pt]
\mathbf{K}_{2,1}(\mathbf{X}_{2},\mathbf{X}_{1}) & 
\mathbf{K}_{2,2}(\mathbf{X}_{2},\mathbf{X}_{2}) & 
\cdots & 
\mathbf{K}_{2,S}(\mathbf{X}_{2},\mathbf{X}_{S}) \\[6pt]
\vdots & \vdots & \ddots & \vdots \\[6pt]
\mathbf{K}_{S,1}(\mathbf{X}_{S},\mathbf{X}_{1}) & 
\mathbf{K}_{S,2}(\mathbf{X}_{S},\mathbf{X}_{2}) & 
\cdots & 
\mathbf{K}_{S,S}(\mathbf{X}_{S},\mathbf{X}_{S})
\end{bmatrix}}
\end{equation}

\textcolor{black}{where the matrix has dimensions $(N_1 + N_2 + ... + N_S) \times (N_1  + N_2 + ... + N_S)$. Alternatively, this can also be viewed as an $S \times S$ block matrix, defined as}
\begin{equation}
\textcolor{black}{\mathbf{K}_{s,s'}(\mathbf{X}_s, \mathbf{X}_{s'}) =
\begin{bmatrix}
k_{s,s'}(\mathbf{x}_1^s, \mathbf{x}_1^{s'}) & k_{s,s'}(\mathbf{x}_1^s, \mathbf{x}_2^{s'}) & \cdots & k_{s,s'}(\mathbf{x}_1^s, \mathbf{x}_{N_{s'}}^{s'}) \\
k_{s,s'}(\mathbf{x}_2^s, \mathbf{x}_1^{s'}) & k_{s,s'}(\mathbf{x}_2^s, \mathbf{x}_2^{s'}) & \cdots & k_{s,s'}(\mathbf{x}_2^s, \mathbf{x}_{N_{s'}}^{s'}) \\
\vdots & \vdots & \ddots & \vdots \\
k_{s,s'}(\mathbf{x}_{N_s}^s, \mathbf{x}_1^{s'}) & k_{s,s'}(\mathbf{x}_{N_s}^s, \mathbf{x}_2^{s'}) & \cdots & k_{s,s'}(\mathbf{x}_{N_s}^s, \mathbf{x}_{N_{s'}}^{s'})
\end{bmatrix} \in \mathbb{R}^{N_s \times N_{s'}}}
\label{equ:form_1_kernel}
\end{equation}

\textcolor{black}{which encodes the cross-covariances between fidelity levels $s$ and $s'$, where $s,s' \in \{1,..,S\}$.}

\textcolor{black}{A common approach used to model $\mathbf{K}_\text{MOGP}$ involves assuming a separable kernel structure using the Intrinsic Coregionalisation Model (ICM) \cite{bonilla2007multi}. The ICM assumes that the covariance between outputs (different fidelity levels) can be expressed as the product of a task covariance matrix and a shared kernel over the input space. Under this assumption, the full multi-output kernel can be written blockwise as}

\begin{equation}
\textcolor{black}{\mathbf{K} =
\begin{bmatrix}
B_{1,1}\mathbf{K}_{(1,1)} & B_{1,2}\mathbf{K}_{(1,2)} & \cdots & B_{1,S}\mathbf{K}_{(1,S)} \\
B_{2,1}\mathbf{K}_{(2,1)} & B_{2,2}\mathbf{K}_{(2,2)} & \cdots & B_{2,S}\mathbf{K}_{(2,S)} \\
\vdots & \vdots & \ddots & \vdots \\
B_{S,1}\mathbf{K}_{(S,1)} & B_{S,2}\mathbf{K}_{(S,2)} & \cdots & B_{S,S}\mathbf{K}_{(S,S)}
\end{bmatrix}}
\end{equation}

\textcolor{black}{where $\textbf{B}\in \mathbb{R}^{S \times S}$ is a positive semi-definite coregionalisation matrix capturing the inter-fidelity correlations, and each $\mathbf{K}_{(s,s')} = [k(\mathbf{x}_i^s, \mathbf{x}_j^{s'})]$ is a standard scalar-valued kernel (e.g., RBF, Matern) evaluated over the input space. }

\textcolor{black}{This type of formulation of a multi-output GP is called a multi-task Gaussian process (MTGP). This can be written as a blockwise Hadamard product, which reshapes the data to index each point by both input and fidelity level, forming $\textbf{K}_{\text{MT}}=\textbf{B}_{(s,s')}\odot \mathbf{K}(\textbf{X}_s, \textbf{X}_{s'})$, which is an element-wise product applied blockwise, and $\mathbf{K}(\textbf{X}_s, \textbf{X}_{s'})$ is a block matrix containing input space covariances $k(\mathbf{x}_i, \mathbf{x}_j)$, structured to align with the fidelity pairs indexed in $\textbf{B}$. This is distinct from the Kronecker-structured MTGP model, which can be more computationally efficient due to the smaller effective covariance inversions. However, it is structurally rigid, requiring that inputs be evaluated at each fidelity level. This may be impractical when high-fidelity evaluations are expensive; as a result, Hadamard-based MTGP has been employed.}

\textcolor{black}{Another popular multi-output BO approach is hierarchical co-kriging \cite{kennedy2000predicting}, which models each higher-fidelity output as a scaled version of the next lower-fidelity plus a residual term used to model discrepancy. This is especially useful when lower fidelity outputs systematically under- or over-estimate the true response, introducing bias. However, preserving this nested structure would involve evaluating a potentially expensive function at each fidelity level, making the process prohibitively expensive and often impractical. Other multi-fidelity surrogates include deep Gaussian processes \cite{damianou2013deep}, which stack GPs in a neural-network-like hierarchy to capture complex nonlinearities. In addition, artificial neural networks (ANNs) can flexibly learn intricate cross-fidelity mappings given a sufficiently large dataset, which provides a more scalable alternative to GPs.}

\section{Detailed derivations and multi-fidelity acquisition formulations}
\label{appendix:Detailed derivations and multi-fidelity acquisition formulations}

\textcolor{black}{(1) Log-EI and its extension, MF-EI}

\textcolor{black}{In single-fidelity BO, the EI can be written as}
\textcolor{black}{\begin{equation}
    \text{EI}(\mathbf{x}) = \sigma_{\text{post}}(\mathbf{x}) \left[ z(\mathbf{x}) \Phi(z(\mathbf{x})) + \phi(z(\mathbf{x})) \right], \quad 
    z(\mathbf{x}) = \frac{\mu(\mathbf{x}) - f^*}{\sigma_{\text{post}}(\mathbf{x})}
\end{equation}}
\textcolor{black}{where $\mu$ and $\sigma_{\textbf{post}}$ denote the GP posterior mean and standard deviation. The Log-EI acquisition is defined as }
\textcolor{black}{\begin{equation}
    \text{Log-EI}(\textbf{x}) = \text{Log}\left[ \text{EI}(\textbf{x}) + \epsilon \right]
\end{equation}}
\textcolor{black}{with $\epsilon > 0$ for stability.}

\textcolor{black}{Following \cite{huang2006sequential}, a multi-fidelity form is obtained by incorporating fidelity-dependent corrections}
\textcolor{black}{\begin{equation}
    \text{MF-Log-EI} = \text{Log-EI}(\textbf{x}) \cdot \text{CR} \cdot \tilde{\rho}(s)
\end{equation}}
\textcolor{black}{where $\text{CR} = C(S)/C(s)$ and $\tilde{\rho}(s)$ is the Pearson correlation between the posterior predictions at fidelities $s$ and $S$}
\textcolor{black}{\begin{equation}
    \tilde{\rho} = \frac{\text{Cov}(f^{s}(\textbf{x}_i),f^{S}(\textbf{x}_i))}{\sqrt{\text{Var}(f^{s}(\textbf{x}_i))\text{Var}(f^{S}(\textbf{x}_i))}} = \frac{k_{s,S}(\textbf{x}_i,\textbf{x}_i)}{\sigma^{s}_{\text{post}(\textbf{x}_i)}\sigma^{S}_{\text{post}(\textbf{x}_i)}}.
\end{equation}}
\textcolor{black}{where $\textbf{x}_i \in \mathcal{X}$ are points sampled from the design space by sampling the Sobol' sequence, used to estimate the predictive mean, variances and covariances.}

\textcolor{black}{(2) MF-UCB}

\textcolor{black}{While MF-Log-EI is a direct extension of the standard single-fidelity Log-EI, modified through the introduction of fidelity-dependent coefficients, MF-UCB does not explicitly facilitate multi-fidelity in such a manner. Instead, it adopts its structural form and redefines its components to accommodate the multi-fidelity setting. Although MF-UCB draws inspiration from the MF-LCB formulation in \cite{jiang2019variable}, which employs separate kriging surrogates for each fidelity, the acquisition function implementation is adapted for use with MTGP. This eliminates the need to introduce scaling factors and hierarchical structure in obtaining components to form the acquisition function. }

\textcolor{black}{The MF-UCB function is defined as}
\textcolor{black}{\begin{equation}
    \text{MF-UCB}(\textbf{x},s) = \omega_1 \cdot \mu^s(\textbf{x})+\omega_2 \cdot \sigma^s_{\text{post}}(\textbf{x}) \cdot \text{CR}
\end{equation}}
\textcolor{black}{where, $\mu^s(\textbf{x})$ and $\sigma^s_{\text{post}}(\textbf{x})$ are the posterior mean and standard deviation of the MTGP model at fidelity $s$, and CR is a cost ratio term. The weights $\omega_1$ and $\omega_2$ are given by}
\textcolor{black}{\begin{equation}
    \omega_1=\frac{a_1}{a_1+a_2+\epsilon}, \quad \omega_2=\frac{a_2}{a_1+a_2+\epsilon}
\end{equation}}
 \textcolor{black}{where $\omega_1 + \omega_2 \approx 1$, with $a_1$ and $a_2$ computed from the highest fidelity model predictions as}
\textcolor{black}{\begin{equation}
    a_1 = \frac{\sqrt{\mathbb{E}[(\mu^S(\mathbf{x}_i)^2]-(\mathbb{E}[\mu^S(\mathbf{x}_i)])^2}}{|\mathbb{E}[\mu^S(\mathbf{x}_i)]|}, \quad  a_2 = \frac{\sqrt{\mathbb{E}[(\sigma^S(\mathbf{x}_i))^2]-(\mathbb{E}[\sigma^S(\mathbf{x}_i)])^2}}{\mathbb{E}[\sigma^S(\mathbf{x}_i)]}
\end{equation}}
\textcolor{black}{Here, $a_1$ and $a_2$ are coefficients of variation that quantify the landscape variability of the highest-fidelity predictions. A large $a_1$, favours exploitation; a large $a_2$ favours exploration. Only the uncertainty term is cost-scaled, encouraging exploratory queries at lower fidelities.}

\textcolor{black}{The method assumes that the highest-fidelity posterior is sufficiently informative and uses ARD kernels to help the GP capture relevant input dimensions.}

\textcolor{black}{(3) Sequential MF-MES}

\textcolor{black}{A sequential acquisition function is also utilised, which starts by selecting the input location, then the fidelity level.}

\textcolor{black}{\emph{Step 1 (location):}}
\textcolor{black}{\begin{equation}
    \textbf{x}_{\text{next}} = \arg\max \text{Log-EI}(\textbf{x}, S)
\end{equation}}
\textcolor{black}{or can be replaced with any other single-fidelity acquisition function.}

\textcolor{black}{\emph{Step 2 (fidelity):}}

\textcolor{black}{The MES acquisition function measures the information gained about the unknown maximum function value $\text{y}^* = f(\textbf{x}_{\text{next}})$. The single-fidelity MES is}
\textcolor{black}{\begin{equation}
    I(\text{y}^*;\text{y}|O,\textbf{x})=H(p(\text{y}|O,\textbf{x}))-\mathbb{E}_{\text{y}^*\sim p(\text{y}^*|O)}\left[H(p( \text{y}|O,\textbf{x},\text{y}^*)) \right].
\end{equation}}
\textcolor{black}{The multi-fidelity variant divides this by the cost}
\textcolor{black}{\begin{equation}
    \text{MF-MES}(\textbf{x}, s) = \frac{\ I(\text{y}^*;\text{y}^s|O,\textbf{x})}{C(s)}.
\end{equation}}
\textcolor{black}{Given $\textbf{x}_{\text{next}}$, the fidelity is chosen as}
\textcolor{black}{\begin{equation}
    s_{\text{next}} = \arg\max\text{MF-MES}(\textbf{x}_{\text{next}}, s).
\end{equation}}
\textcolor{black}{This sequential approach has been used in recent MFBO studies \cite{irshad2024leveraging, folch2023combining}. Although Step 1 optimises the acquisition at the highest fidelity, the MTGP ensures that its predictions, and by extension, the acquisition values are informed by correlations with low-fidelity data.}

\section{Additional details of MTGP and BO methodology}
\label{appendix:Additional details of MTGP and BO methodology}

\textcolor{black}{In addition to evaluating correlation coefficients, the impact of resolution and design parameters $\boldsymbol{\theta}$ on simulation time was also examined. The structures shown in Fig.~\ref{fig:spinodoids} were generated at three different resolutions and subsequently crushed. For each simulation, the total wall time, measured from structure generation to the writing of the output file, was recorded. Results for the four cases are presented in Fig.~\ref{fig:initial_data_sim_time_correlation}(d). While a general trend between resolution and simulation time may appear trivial, the simulation time varied significantly across different structures, suggesting that design parameters have a notable influence on computational cost.}

\textcolor{black}{Although absolute simulation times may differ across hardware setups, the consistency of hardware used throughout data acquisition and optimisation ensures that the recorded times are representative for this study. As such, simulation time can be interpreted as the computational cost of evaluating the point obtained from the acquisition function, now dependent on both the resolution and design parameters, expressed as $C(\boldsymbol{\theta},s)$. To streamline the optimisation process, this cost can be averaged across data points at different fidelity levels, but due to its dependence on $\boldsymbol{\theta}$, an additional surrogate model is required. In this work, a simple single-output GP regressor is employed to fit the computational cost. While an MTGP could be used, the added computational expense does not justify its benefits in this context. }

\textcolor{black}{The initial dataset was generated using a simulation time budget of 30,000 seconds, during which individual simulation times were logged per evaluation. Input selection during this phase was based on fidelity-weighted sampling, giving preference to lower-fidelity evaluations to sufficiently map the design space. The total evaluation budget for each optimisation run was set to 200,000 seconds. This was empirically chosen to provide sufficient computational resources for meaningful convergence, while also enabling comparisons across methods. Larger budgets can be specified if higher reconstruction accuracy is required. In addition to the simulation time budget, an iteration-based stopping criterion was implemented to prevent excessive reliance on low-cost, low-fidelity evaluations, especially relevant when acquisition functions aggressively favour cheaper approximations. These dual stopping criteria are particularly important because GPs in general scale poorly with increasing dataset size, with a time complexity of $\mathcal{O}(N^3)$ \cite{liu2020gaussian}. This computational burden is further amplified in MTGPs, where the inclusion of additional tasks increases both model complexity and training cost. In this context, it is assumed that candidate generation is computationally negligible compared to the cost of evaluating those candidates. As a result, the total simulation time is taken as a proxy for the overall optimisation wall time. This assumption is generally valid in scenarios where candidate evaluations (e.g., via high-fidelity simulations) are significantly more time-consuming than optimising the acquisition function used. However, this assumption breaks down with larger datasets, where optimising the acquisition function may require multiple restarts or additional iterations to achieve convergence to find suitable candidates. }

\textcolor{black}{All optimisation routines in this study were implemented using BoTorch \cite{balandat2020botorch}, a flexible and modular library for BO built on PyTorch. The library natively supports automatic differentiation, and GPU acceleration allows for efficient optimisation of GP parameters and scalable evaluations of acquisition functions. To accommodate the requirements of this study, custom implementations of various MFBO methods were developed within the BoTorch framework. The framework supports asynchronous evaluation of candidate points through parallel batch generation; in this work, a batch size of 2 was chosen. This configuration facilitated the efficient execution of low-fidelity simulations while avoiding significant bottlenecks during concurrent high-fidelity evaluations. The optimisation was conducted on the cluster using 6 cores with 20 GB RAM per core. This level of computational resource was necessary to support the MF-MES method, due to its high memory requirements for entropy estimation and optimisation of dual acquisition functions. In contrast, such resources were not required for MF-UCB, MF-EI, or single-fidelity optimisation, which are significantly less memory-intensive.}

\section{Further information on synthetic function benchmarking and final recommendation for evaluation}
\label{appendix:Synthetic function benchmarking}

The two synthetic benchmark functions were:

(1) \emph{Augmented Branin}

This is a 3-dimensional function, where the last dimension, $s$, represents the fidelity, \\ 
$(x_1, x_2, s) \in [-5,10] \times [0,15] \times[0,1]$.
The function is given by:
\begin{align}
B(x_1,x_2,s)=\ &
\bigl(x_2 - (\,b - 0.1(1-s)\,)x_1^{2} + c x_1 - r\bigr)^{2} \notag\\
&\quad + 10(1-t)\cos(x_1) + 10
\end{align}
where the constants are
\begin{equation}
    b=\frac{5.1}{4\pi^2}, \quad c=\frac{5}{\pi}, \quad r=6, \quad t=\frac{1}{8\pi}
\end{equation}

For this benchmark function, a budget of 4 units was used per fidelity; hence, 12 units in total were used to generate the initial data, resulting in 4, 8, and 40 evaluated points, ranging from high to low-fidelity, respectively. Generating data using this method allows the dataset to be consistently sampled with evaluations at all fidelities present. This method enables consistent sampling of the dataset, ensuring accurate posterior predictions. The optimisation was terminated using a budget-based stopping criterion; it stopped once the total budget reached 30 units. This particular benchmark has multiple global optima at various fidelity levels; hence, the final recommendation was obtained based solely on which set of input parameters yielded the highest objective value within the optimisation budget.

(2) \emph{Augmented Hartmann}

This function is a 7-dimensional synthetic benchmark defined over the domain 
\newline 
$(x_1, x_2, x_3, x_4, x_5, x_6, s) \in [0, 1]^7$ where the last dimension $s$ serves as the fidelity parameter. It is defined as
\begin{align}
H(\mathbf{x}) =\ & -\left( \alpha_1 - 0.1(1 - x_7) \right) 
\exp\left( - \sum_{j=1}^6 A_{1j}(x_j - P_{1j})^2 \right) \notag \\
&\quad - \sum_{i=2}^4 \alpha_i \exp\left( - \sum_{j=1}^6 A_{ij}(x_j - P_{ij})^2 \right)
\end{align}
where $\boldsymbol{\alpha} = [1.0,\ 1.2,\ 3.0,\ 3.2]$ , $A \in \mathbb{R}^{4 \times 6}  $and  $ P \in \mathbb{R}^{4 \times 6} $ are defined as

\begin{equation}
    A = \begin{bmatrix}
10 & 3 & 17 & 3.5 & 1.7 & 8 \\
0.05 & 10 & 17 & 0.1 & 8 & 14 \\
3 & 3.5 & 1.7 & 10 & 17 & 8 \\
17 & 8 & 0.05 & 10 & 0.1 & 14
\end{bmatrix}
\end{equation}
\begin{equation}
    P = \frac{1}{10^4} \cdot \begin{bmatrix}
1312 & 1696 & 5569 & 124 & 8283 & 5886 \\
2329 & 4135 & 8307 & 3736 & 1004 & 9991 \\
2348 & 1451 & 3522 & 2883 & 3047 & 6650 \\
4047 & 8828 & 8732 & 5743 & 1091 & 381
\end{bmatrix}
\end{equation}

An evaluation budget of 8 units was used for this benchmark due to its increased input dimensionality. This resulted in a dataset with 8 samples from high-fidelity, 16 from medium-fidelity and 80 from lowest fidelity. For this benchmark, a total budget of 60 units was used as the stopping criterion, again due to the increased dimensionality. The optimum for this benchmark can only be identified at the highest fidelity level. Consequently, determining the final recommendation of input parameters that yields the highest objective value requires a more involved process. Specifically, a set of input recommendations is obtained at fidelity levels $s \in \{1, 2, 3\}$. These three candidates are then evaluated using the highest-fidelity function, and the final recommendation is selected as 
\begin{equation}
    \text{y}_\text{rec}=\max_{s \in \{1, 2, 3\}}f^S(\mathbf{x}_{\text{rec}}^s)
\end{equation}
where $\mathbf{x}_{\text{rec}}^s$ is the recommended input from fidelity s. The input yielding the highest objective value is ultimately chosen as the optimum. It should be mentioned that the MF-MES method was omitted for this benchmark due to its high computational cost, since it involves optimising multiple acquisition functions, making it significantly more expensive than other approaches, especially for this high-input-dimension problem.

{Nonetheless, benchmarking on synthetic functions was carried out to evaluate the initial performance of the proposed methods and to verify the framework. Two multi-fidelity benchmark functions were employed to establish baseline performance for the strategies described in Subsection.~\ref{subsection: Multi-fidelity Bayesian Optimisation}. An initial dataset was generated across three discrete fidelity levels $s \in \{1, 2, 3\}$, ranging from low to high fidelity by sampling the Sobol' sequence. Each discrete level corresponds to a fixed value of 0.1, 0.5, and 1, respectively, which is the last dimension of each function. The cost of evaluation at a given fidelity level was determined by the function $C(s) = s$, resulting in a linear scaling with fidelity. The optimisation process was repeated five times with different seeds. For both of the benchmarks, a batch size of 4 was utilised to allow multiple, parallel evaluations of the function to speed up the overall optimisation cycle. }

\begin{table}[h]
{
\centering
\begin{tabular}{@{}ccccccc@{}}
\toprule
Function & \shortstack{True\\Optimum} & Stat & \shortstack{Single\\Fidelity} & MF-EI & MF-UCB & MF-MES \\
\midrule
\multirow{3}{*}{\makecell[c]{Augmented\\Branin}} 
  & \multirow{3}{*}{\makecell[c]{-0.3979}} & Q1 & -0.4485 & -0.3999 & -0.4407 & -0.4611 \\
  &                                       & Q2 & -0.4173 & -0.3985 & -0.4074 & -0.4182 \\
  &                                       & Q3 & -0.4109 & -0.3981 & -0.4019 & -0.3999 \\
\addlinespace
\multirow{3}{*}{\makecell[c]{Augmented\\Hartmann}} 
  & \multirow{3}{*}{\makecell[c]{3.3224}}  & Q1 & 2.4627  & 3.1993  & 2.8321 & -- \\
  &                                       & Q2 & 3.01  & 3.3102  & 3.2031  & -- \\
  &                                       & Q3 & 3.2064  & 3.3188  & 3.3214  & -- \\
\bottomrule
\end{tabular}
\caption{\label{table:benchmark optimal values}{Comparison of optimisation methods across benchmark functions. Q1, Q2 (median), and Q3 are reported for each method.}}}
\end{table}

{Table.~\ref{table:benchmark optimal values} presents the optimal values identified by the four methods with the performance of each method evaluated using the Q1, Q2, and Q3 metrics. These results indicate that the multi-fidelity methods achieve performance compared to, or even better than, the single-fidelity approach, which relies exclusively on high-fidelity evaluations. As shown in Table.~\ref{table:benchmark optimal values}, for the \emph{Augmented Branin} function, MF-EI, consistently achieves the lowest difference (between optimal and obtained values) across Q1, Q2, and Q3. This indicates both accuracy and robustness across different initialisations. The single-fidelity method underperforms relative to MF-EI and MF-UCB, particularly in the lower quartile, reflecting less stable convergence. For the \emph{Augmented Hartmann} function, MF-EI again shows superior results, with a Q3 value of 3.3188, nearly matching the optimal value of 3.3224. The narrow spread between Q1 and Q3 suggests reliable optimisation across different Sobol' sequences. Meanwhile, the single-fidelity method displays greater variability, with a significantly lower Q1 value, highlighting inconsistent performance. Overall, the table highlights that multi-fidelity approaches, especially MF-EI, can deliver both accurate and stable optimisation performance while offering a more cost-effective strategy than relying solely on high-fidelity evaluations.}

\begin{figure}[ht]
    \centering
    \includegraphics[width=1\textwidth]{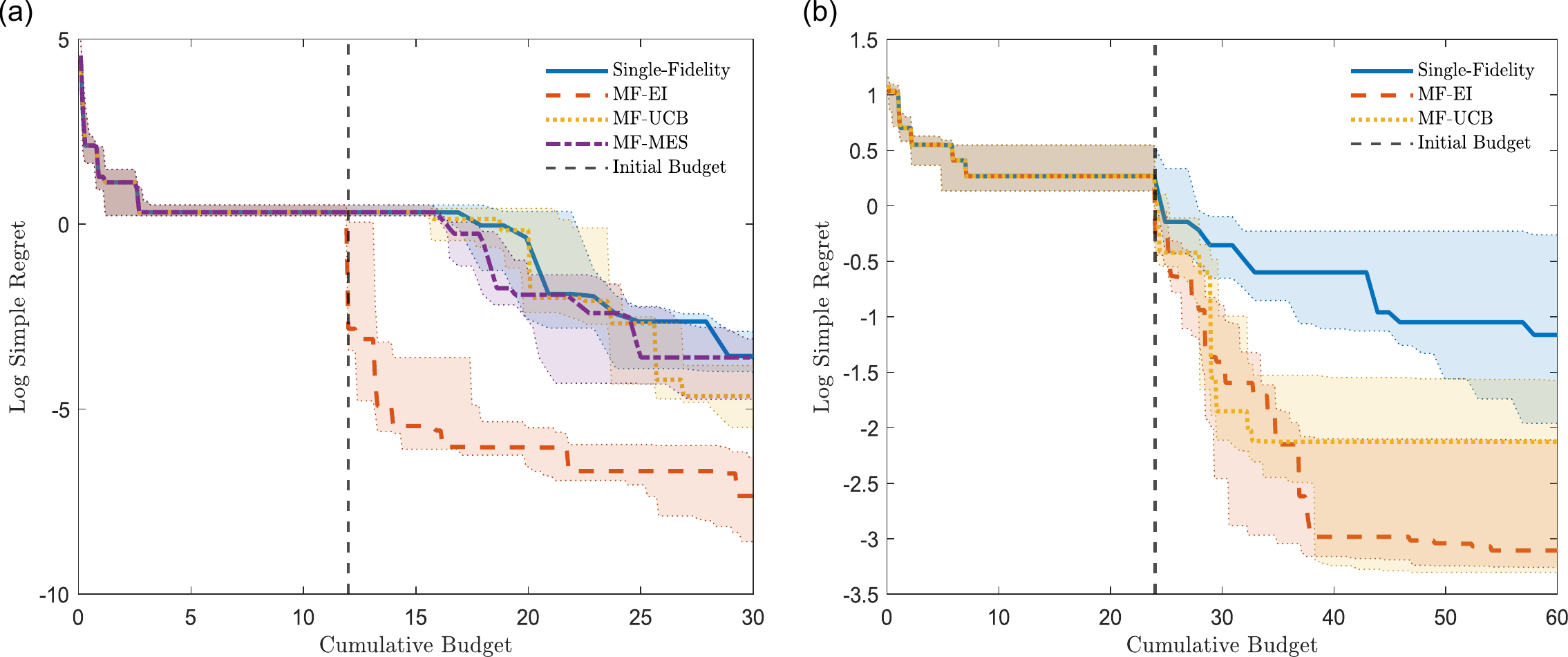}
    \caption[Cumulative history plots for two synthetic multi-fidelity Bayesian optimisation benchmark functions (a) \emph{Augmented Branin}, and (b) \emph{Augmented Hartmann}.]{{Cumulative history plots for two synthetic multi-fidelity Bayesian optimisation benchmark functions (a) \emph{Augmented Branin}, and (b) \emph{Augmented Hartmann}. For each method, the first quartile (Q1), median (Q2), and third quartile (Q3) are shown across five independent optimisation runs. Initial data for each run was generated using a Sobol' sequence with a different random seed.}}
    \label{fig:MFBO_benchmarks}
\end{figure}

{The cumulative optimisation history plot shown in Fig.~\ref{fig:MFBO_benchmarks}, illustrates the convergence behaviour of each optimisation method over the evaluation budget, where the objective is measured using simple regret. For each method, the bold lines with varying linestyles represent the median, and the shaded regions indicate the interquartile range (IQR). This allows a direct comparison of how quickly and consistently each method approaches the true optimum. From the plot, it is evident that the multi-fidelity methods, especially MF-EI, tend to achieve faster convergence and lower final regret compared to the single-fidelity baseline in both benchmarks. Additionally, MF-EI shows the most rapid initial improvement, often outperforming the other methods within the early stages of optimisation. This is also true for MF-UCB for the \emph{Augmented Hartmann} benchmark, but reaches a plateau quickly, which is possibly an indication of preference for exploration over exploitation. However, the best run for MF-UCB outperforms MF-EI for this particular benchmark. However, for the \emph{Augmented Branin} benchmark function, the convergence for single-fidelity, MF-UCB, and MF-MES follows a similar convergence trend, but are vastly outperformed by the MF-EI method. Regardless, the best runs of MF-UCB and MF-MES still outperform the single-fidelity method.}

{Importantly, the benchmarks also reveal that the composition of the initial dataset has a significant impact on performance. This is especially evident in the \emph{Augmented Hartmann} benchmark function, which is a higher-dimensional function in comparison to the \emph{Augmented Branin}. As a result, the initial dataset becomes comparatively sparse, providing weaker coverage of the design space and less information for establishing cross-fidelity correlations. Consequently, the optimisation becomes more sensitive to the specific Sobol’ realisation used to generate the initial points, leading to greater run-to-run variability. This effect is reflected in the widening interquartile ranges as additional budget is expended, where each trial tends to diverge more noticeably despite following the same acquisition policy. The benchmarks, therefore, highlight that, particularly in higher-dimensional settings, a sufficiently rich and well-distributed initial multi-fidelity dataset is crucial for stable convergence and robust optimisation performance. Nevertheless, even in this high-dimensional setting, it remains advantageous to use MF-EI, as its acquisition function is relatively inexpensive to optimise when selecting the next set of input points, unlike MF-MES, whose acquisition computation is considerably more costly. Therefore, a well-structured initial multi-fidelity dataset combined with a computationally efficient acquisition function like MF-EI ensures both robust convergence and practical tractability, even in higher-dimensional problems.}

\section{Full inverse design results}
\label{appendix:Optimisation history MFBO}
Each subsection presents the complete optimisation results corresponding to a specific target mechanical response. For each optimisation method, the best-performing result is shown through a table with the sets of $\boldsymbol{\theta}$ values and the corresponding similarity score obtained from the recommendation strategy outlined in Sec.~\ref{sec:results MFBO}. In addition, the mechanical response curves are also shown in a separate figure obtained at the recommended inputs, which have been evaluated at the highest fidelity. This is accompanied by three additional plots that display the design recommendations obtained at each fidelity level for all multi-fidelity approaches used. This is followed by an optimisation history plot, providing a pointwise comparison of how the similarity score evolves throughout the optimisation, measured in terms of cumulative cost. Additionally, plots of cumulative maxima achieved within a given computational budget are included for each target case. Finally, a summary is provided of the number of evaluations performed by each method, along with a breakdown of how these evaluations are distributed across fidelity levels. 

\subsection{Target 2}

Table.~\ref{table:target_2_recommendations} summarises the optimisation results for Target 2 as the desired mechanical response. All methods yield geometries closely matching the target, typically with $\theta_3 \approx 50^{\circ}$ and low $\theta_1$ and $\theta_2$ values. Multi-fidelity methods outperform the single-fidelity approach, with best-case averaging 35\% higher similarity scores. Among these, MF-UCB achieves the highest score.

Fig.~\ref{fig:target_2_combined_results} shows the mechanical responses from each method, with the target in solid red; all closely follow the desired response. The optimisation histories in Fig.~\ref{fig:target_2_optimisation_history}(a-d) reveal distinct search behaviours: Single-fidelity and MF-EI converge rapidly with limited exploration, whereas MF-UCB and MF-MES explore more extensively. MF-UCB predominantly evaluates at the lowest fidelity, while MF-MES favours resolution 25.

 From Fig.~\ref{fig:target_2_optimisation_history}(e), all multi-fidelity methods reach their peak similarity faster than single-fidelity. Although MF-MES identified an input with the highest overall similarity (across all fidelities), its performance dropped when this point was re-evaluated at the highest fidelity.
 
 Finally, the stacked bar chart in Fig.~\ref{fig:target_2_optimisation_history}(f) shows a pattern similar to Target 1: MF-UCB mainly uses resolution 20 (yielding the most evaluations), followed by MF-MES at resolution 25, then MF-EI splitting effort between resolutions 25 and 30, with single-fidelity naturally having the fewest evaluations.

\begin{table}[]
\centering
\renewcommand{\arraystretch}{0.6}
\begin{tabular}{@{}cccccc@{}}
\toprule
Method & Rec. \# & $\theta_1$ & $\theta_2$ & $\theta_3$ & $\mathcal{S}$\\
\midrule
\textbf{Target 2} & -- & \textbf{7.65} & \textbf{5.69} & \textbf{50.39} & -- \\

\addlinespace
Single-Fidelity & 1 & 4.44  &  2.06 & 55.1  &  -0.00047 \\

\addlinespace
\multirow{3}{*}{MF-EI} 
  & 1 & 0 & 11.15 & 50.16 & -0.00033 \\
  & 2 & 16.14 & 1.34 & 46.06 & -0.00056\\
  & 3 & 0 & 17.95 & 50.97 & -0.0011\\

\addlinespace
\multirow{3}{*}{MF-UCB} 
  & 1 & 0.37 & 0.043  & 54.53 &  -0.00028\\
  & 2 & 9.52 & 3.98 & 47.45 &  -0.00032\\
  & 3 & 7.67 & 0.46 & 54.64 &  -0.00074\\

\addlinespace
\multirow{3}{*}{MF-MES} 
  & 1 & 14.98 & 0 & 48.79 & -0.0003 \\
  & 2 & 0 & 8.24 & 53.63 &  -0.00057\\
  & 3 & 11.1 & 0 & 56.8 & -0.0032 \\

\bottomrule
\end{tabular}
\caption{\label{table:target_2_recommendations}Target 2: Comparison of recommended $\theta$'s obtained from each optimisation method, and similarity score $\mathcal{S}$ with respect to the target mechanical response, evaluated at highest fidelity.}
\end{table}

\begin{figure}[]
    \centering
    \includegraphics[width=1\textwidth]{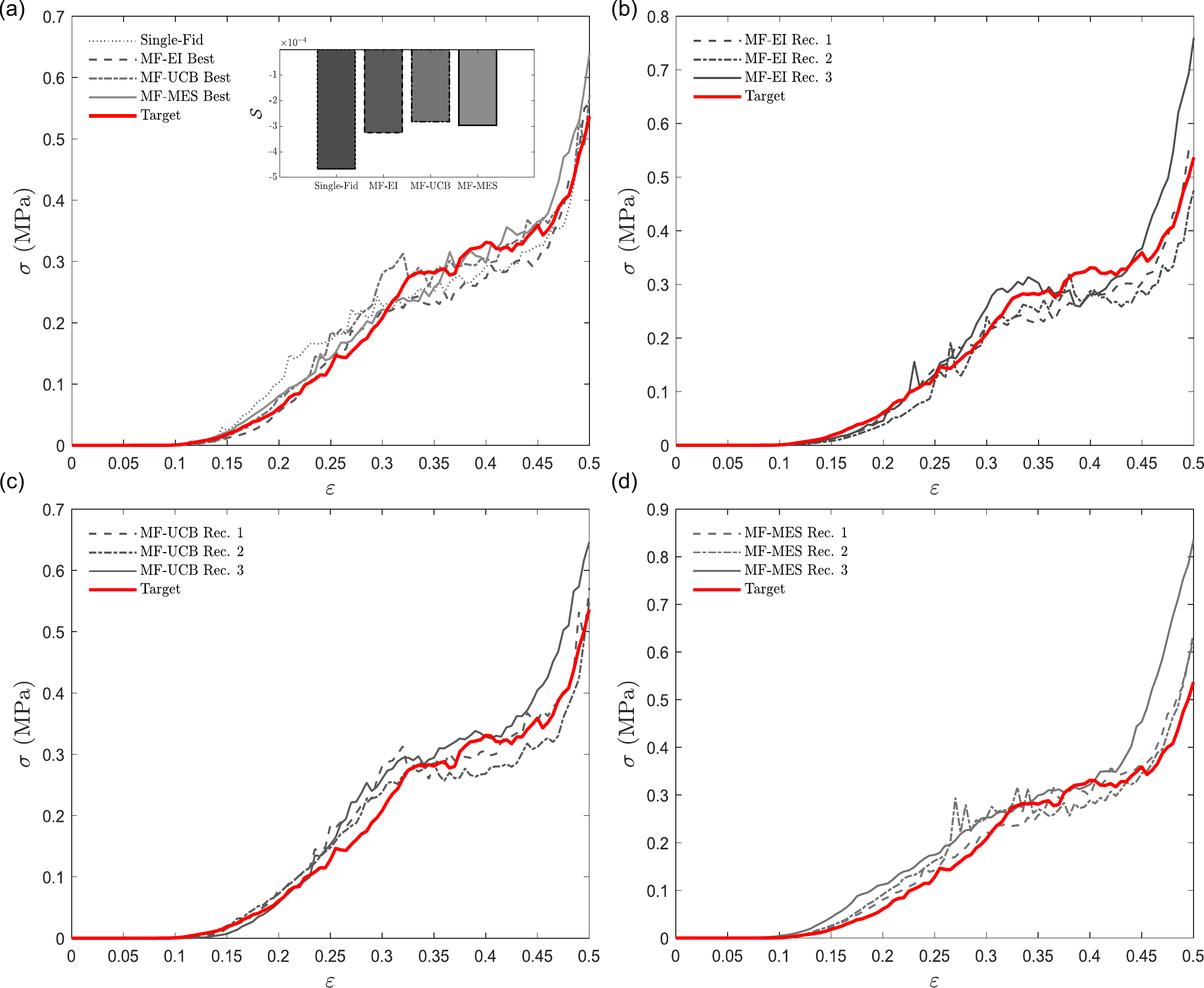}
    \caption{Target 2: (a) Results with the highest similarity score from each method. (b-d) Comparison between the target response and the top three input configurations, based on similarity score, obtained using (b) MF-EI, (c) MF-UCB, and (d) MF-MES.}
    \label{fig:target_2_combined_results}
\end{figure}

\begin{figure}[]
    \centering
    \includegraphics[width=0.9\textwidth]{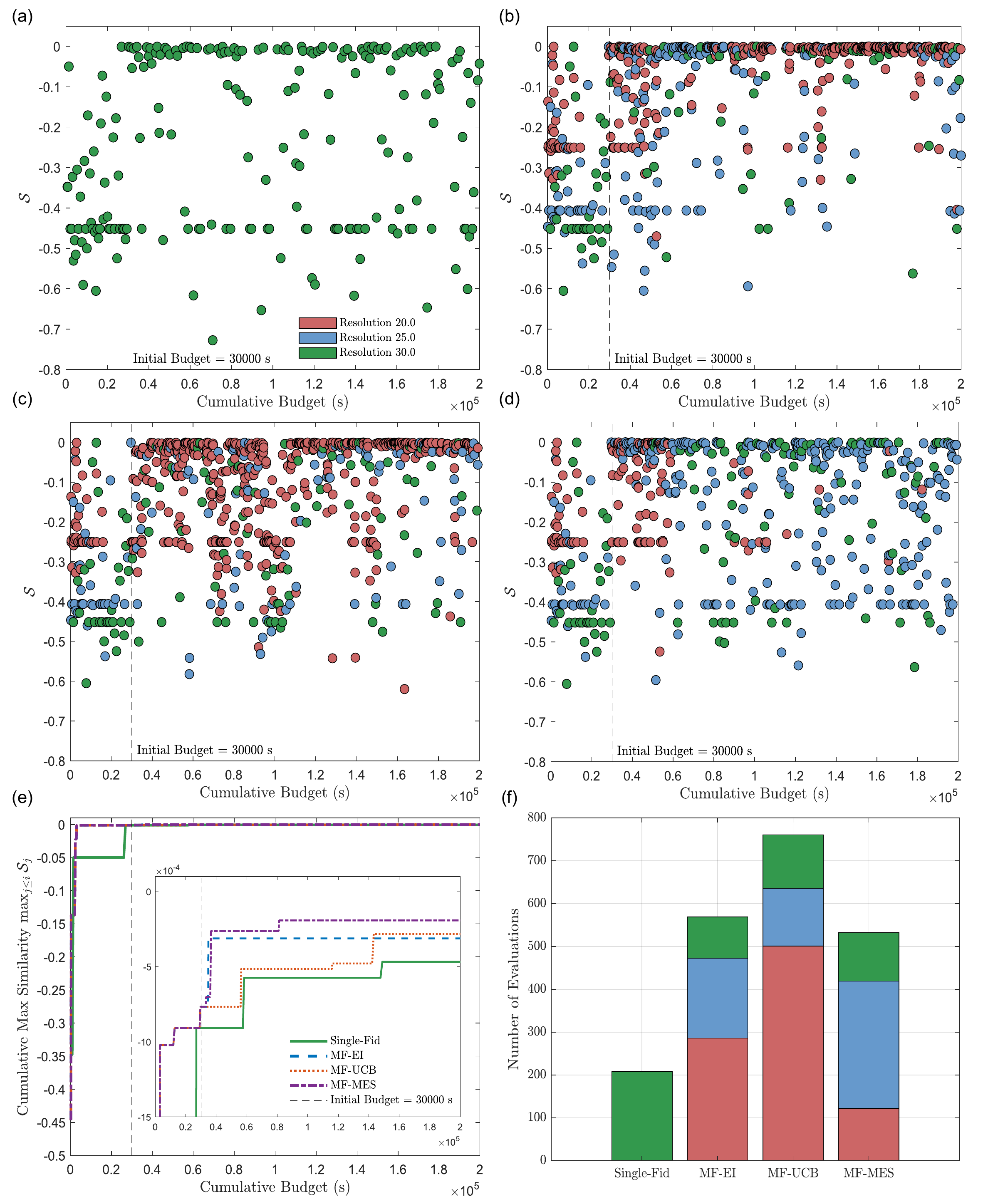}
    \caption[Target 2: Evolution of similarity scores and fidelity usage across four optimisation methods.]{Target 2: Evolution of similarity scores and fidelity usage across four optimisation methods. (a-d) Similarity score $\mathcal{S}$ plotted against cumulative computational budget (in seconds) for each method: (a) Single-fidelity, (b) MF-EI, (c) MF-UCB, and (d) MF-MES. (e) Comparison of cumulative maximum similarity scores under budget constraints, with inset magnifying the overlapping region. (f) Stacked bar chart showing evaluation counts per fidelity for all methods, highlighting resource allocation.}
\label{fig:target_2_optimisation_history}
\end{figure}

\subsection{Target 3}

The optimal set of parameters yielding a mechanical response most similar to Target 3 is presented in Table.~\ref{table:target_3_recommendations}. Consistent with the inverse design results for Targets 1 and 2, the multi-fidelity methods outperformed the single-fidelity inverse design, with the best cases showing a 32\% improvement in similarity scores on average. Among these, MF-MES achieved the highest similarity score, similar to its performance when inverse designing the response of Target 1. However, following the three fidelity recommendations method, the parameters that yielded the maximum similarity scores across fidelity levels, when evaluated at the highest fidelity, showed vastly differing stress-strain curves. This is a recurrence, as similar behaviour was also observed when optimising for Target 1 response. This could primarily be attributed to the lack of input evaluations at the lowest resolution level. On the other hand, the least amount of fluctuation across the recommendations was obtained from MF-UCB, which primarily utilises resolution 20 evaluations. Nonetheless, as in the case of Target 1, the symmetry of the design space led to recommended inputs where $\theta_2 > \theta_1$, despite the true mechanical input corresponding to $\theta_1 > \theta_2$, as a result, exactly one of the recommended inputs is required to be either $\theta_1 \approx 15^{\circ}$ or $\theta_2 \approx 15^{\circ}$. 

\begin{table}[]
\centering
\renewcommand{\arraystretch}{0.6}
\begin{tabular}{@{}cccccc@{}}
\toprule
Method & Rec. \# & $\theta_1$ & $\theta_2$ & $\theta_3$ & $\mathcal{S}$\\
\midrule
\textbf{Target 3} & -- & \textbf{15} & \textbf{5} & \textbf{0} & -- \\

\addlinespace
Single-Fidelity & 1 & 13.33 & 5.09  &  2.6 &  -0.014  \\

\addlinespace
\multirow{3}{*}{MF-EI} 
  & 1 & 0 & 19.28 & 0 & -0.011 \\
  & 2 & 0 & 12.37 & 0 & -0.017 \\
  & 3 & 2.91 & 15.19 & 0.56 & -0.023 \\

\addlinespace
\multirow{3}{*}{MF-UCB} 
  & 1 & 0 & 16.92 & 0 & -0.0091 \\
  & 2 & 0 & 18.62 & 0 & -0.011 \\
  & 3 & 4.47 & 15.14 & 0 & -0.021 \\

\addlinespace
\multirow{3}{*}{MF-MES} 
  & 1 & 10.3 & 20.1 & 0 & -0.0086 \\
  & 2 & 0 & 14.91 & 0 & -0.018 \\
  & 3 & 0 & 33.02 & 0 & -0.086 \\

\bottomrule
\end{tabular}
\caption{\label{table:target_3_recommendations}Target 3: Comparison of recommended $\theta$'s obtained from each optimisation method, along with their similarity score $\mathcal{S}$ with respect to the target mechanical response, evaluated at highest fidelity.}
\end{table}

The mechanical responses obtained from the recommended input parameters are shown in Fig.~\ref{fig:target_3_combined_results}. Overall, all methods reproduce the initial peak followed by a stress drop; however, not all can capture the full magnitude of the peak stress, though they generally replicate the subsequent damage progression. The pointwise optimisation histories are shown in Fig.~\ref{fig:target_3_optimisation_history}(a-d). The single-fidelity method exhibits a mixture of exploitation and exploration, evident from objective values oscillating between two extremes. A similar pattern appears in MF-EI, but as the evaluation budget diminishes, it converges prematurely to a local minimum, which is indicated by the absence of scattered values, likely due to overconfidence in surrogate predictions. The behaviour of the MF-UCB method illustrated in Fig.~\ref{fig:target_3_optimisation_history}(c) shows an initial increase in the objective value, reflecting the inherently greedy nature of the acquisition function. This acquisition function balances exploration and exploitation, as indicated by the fluctuating similarity scores during the inverse design process. A notable shift from exploration to exploitation is observed when the acquisition function reduces the number of recommendations at lower fidelity values to higher fidelities towards the tail end of the process. This is not the case for the MF-MES method depicted in Fig.~\ref{fig:target_3_optimisation_history}(d), which primarily recommends evaluations at the intermediate fidelity level.

Fig.~\ref{fig:target_3_optimisation_history}(e) shows the maximum similarity score within discrete budget intervals. The trend observed in the previous two targets persists: multi-fidelity methods reach convergence earlier than single-fidelity optimisation. Similar to Target 2, some high similarity scores were obtained at lower fidelities, but re-evaluations at the target fidelity revealed lower scores than initially evaluated. Fig.~\ref{fig:target_3_optimisation_history}(f) presents a breakdown of the number of evaluations per method and fidelity level. As in previous cases, MF-UCB relied most heavily on the lowest resolution, whereas MF-MES predominantly used the resolution of 25. MF-MES also accumulated the greatest total number of evaluations due to its limited use of the highest fidelity. MF-EI adopted a more balanced approach across fidelities, with a preference for resolution 20 and the least usage of resolution 25. The single-fidelity method performed the fewest total evaluations, as expected.

\begin{figure}[]
    \centering
    \includegraphics[width=1\textwidth]{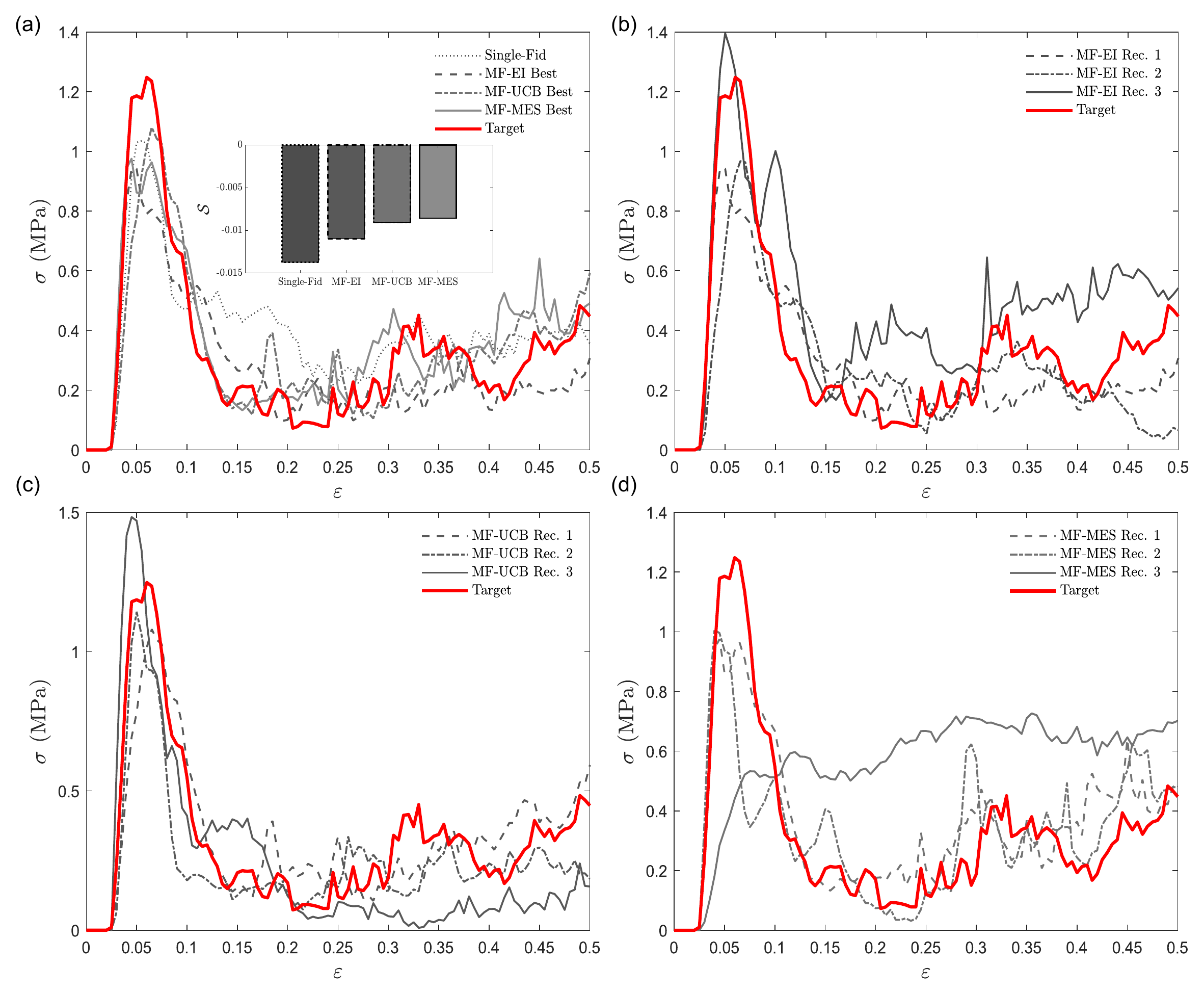}
    \caption{Target 3: (a) Results with the highest similarity score from each method. (b-d) Comparison between the target response and the top three input configurations, based on similarity score, obtained using (b) MF-EI, (c) MF-UCB, and (d) MF-MES.}
    \label{fig:target_3_combined_results}
\end{figure}

\begin{figure}[]
    \centering
    \includegraphics[width=0.9\textwidth]{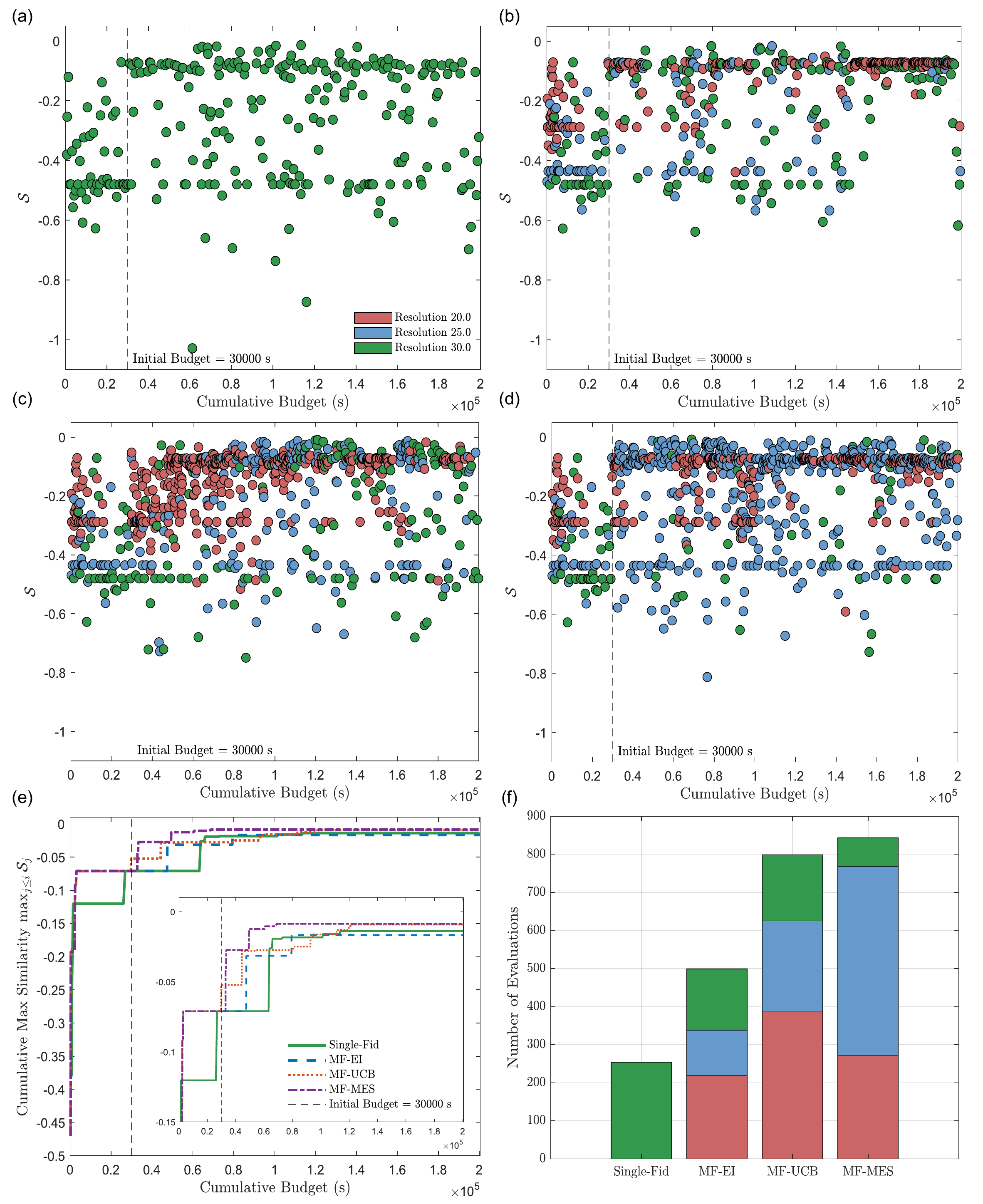}
    \caption[Target 3: Evolution of similarity scores and fidelity usage across four optimisation methods.]{Target 3: Similarity score evolution and fidelity usage for four optimisation methods. (a–d) $\mathcal{S}$ vs.~cumulative computational budget for (a) Single-fidelity, (b) MF-EI, (c) MF-UCB, and (d) MF-MES, with marker colours indicating evaluation fidelity. (e) Cumulative maximum $\mathcal{S}$ comparison under budget constraints, with inset magnifying the overlap region. (f) Stacked bar chart of evaluation counts per fidelity, showing resource allocation across methods.}
    \label{fig:target_3_optimisation_history}
\end{figure}

\subsection{Target 4}

The final target response reflects the behaviour of an ideal energy-absorbing structure, which initially absorbs little energy in the linear-elastic region and then transitions into a long, stable plateau. This plateau, maintained at nearly constant load, arises from mechanisms such as buckling, yielding, or progressive crushing \cite{gibson2003cellular}. Following this definition, the optimal parameters obtained by maximising the similarity score for each method are presented in Table.~\ref{table:target_4_recommendations}. High similarity scores are achieved across all methods, confirming the effectiveness of both single-fidelity and multi-fidelity approaches in reproducing mechanical responses consistent with the target stress-strain curve. Moreover, the multi-fidelity methods consistently outperform the single-fidelity method, achieving across that are on average approximately 37\% higher. Among them, the MF-UCB method delivers the highest similarity score in its best-case recommendation.

\begin{table}[]
\centering
\renewcommand{\arraystretch}{0.6}
\begin{tabular}{@{}cccccc@{}}
\toprule
Method & Rec. \# & $\theta_1$ & $\theta_2$ & $\theta_3$ & $\mathcal{S}$\\
\midrule
\textbf{Target 4} & -- & -- & -- & -- & -- \\

\addlinespace
Single-Fidelity & 1 & 24.99 & 0 & 10.39 & -0.015 \\

\addlinespace
\multirow{3}{*}{MF-EI} 
  & 1 & 25.92 & 0 & 0 & -0.0097 \\
  & 2 & 25.92 & 0 & 0 & -0.01 \\
  & 3 & 37.05 & 8.26 & 0.06 & -0.014 \\

\addlinespace
\multirow{3}{*}{MF-UCB} 
  & 1 & 24.16 & 6.07 & 0.017 & -0.0089 \\
  & 2 & 0 & 23.12 & 0 & -0.012 \\
  & 3 & 12.56 & 28.39 & 8.85 & -0.096 \\

\addlinespace
\multirow{3}{*}{MF-MES} 
  & 1 & 23.42 & 7.82 & 0 & -0.0097 \\
  & 2 & 28.16 & 0 & 0 & -0.012 \\
  & 3 & 26.08 & 5.6 & 0 & -0.014 \\
\bottomrule
\end{tabular}
\caption{\label{table:target_4_recommendations}Target 4: Comparison of recommended $\theta$'s obtained from each optimisation method, along with their similarity score $\mathcal{S}$ with respect to the target mechanical response, evaluated at highest fidelity.}
\end{table}

The mechanical responses obtained from the recommended input parameters are shown in Fig.~\ref{fig:target_4_combined_results}, demonstrating close agreement between the desired response and those generated through the inverse design process. In most cases, the structures exhibited an initial increase in stiffness and a load drop, followed by stress fluctuations oscillating around 0.5 MPa, which is the target plateau. This confirms that the inverse design framework is capable of reproducing arbitrary mechanical behaviour, even when such responses are not explicitly represented within the design space.

Fig.~\ref{fig:target_4_optimisation_history}(a-d) presents the point-wise optimisation history for target response 4. Across all four methods, the histories reflect a balance in policies, fluctuating between exploration and exploitation, with rapid increases in similarity score observed at the early stages of the process. Fig.~\ref{fig:target_4_optimisation_history}(e) shows the cumulative maxima at each budget interval, used to assess convergence. For this target response, the multi-fidelity methods once again converged the fastest, consistently achieving maximum values that surpass those obtained with the single-fidelity method. Finally, Fig.~\ref{fig:target_4_optimisation_history}(f) provides a breakdown of the number of evaluations at each fidelity level. As in previous cases, MF-UCB employed the greatest number of evaluations, with the highest concentration at resolution 20. MF-MES also made extensive use of resolution 25, while MF-EI relied on a broader mix of fidelities, including a relatively large number of high-fidelity evaluations.

\begin{figure}[]
    \centering
    \includegraphics[width=1\textwidth]{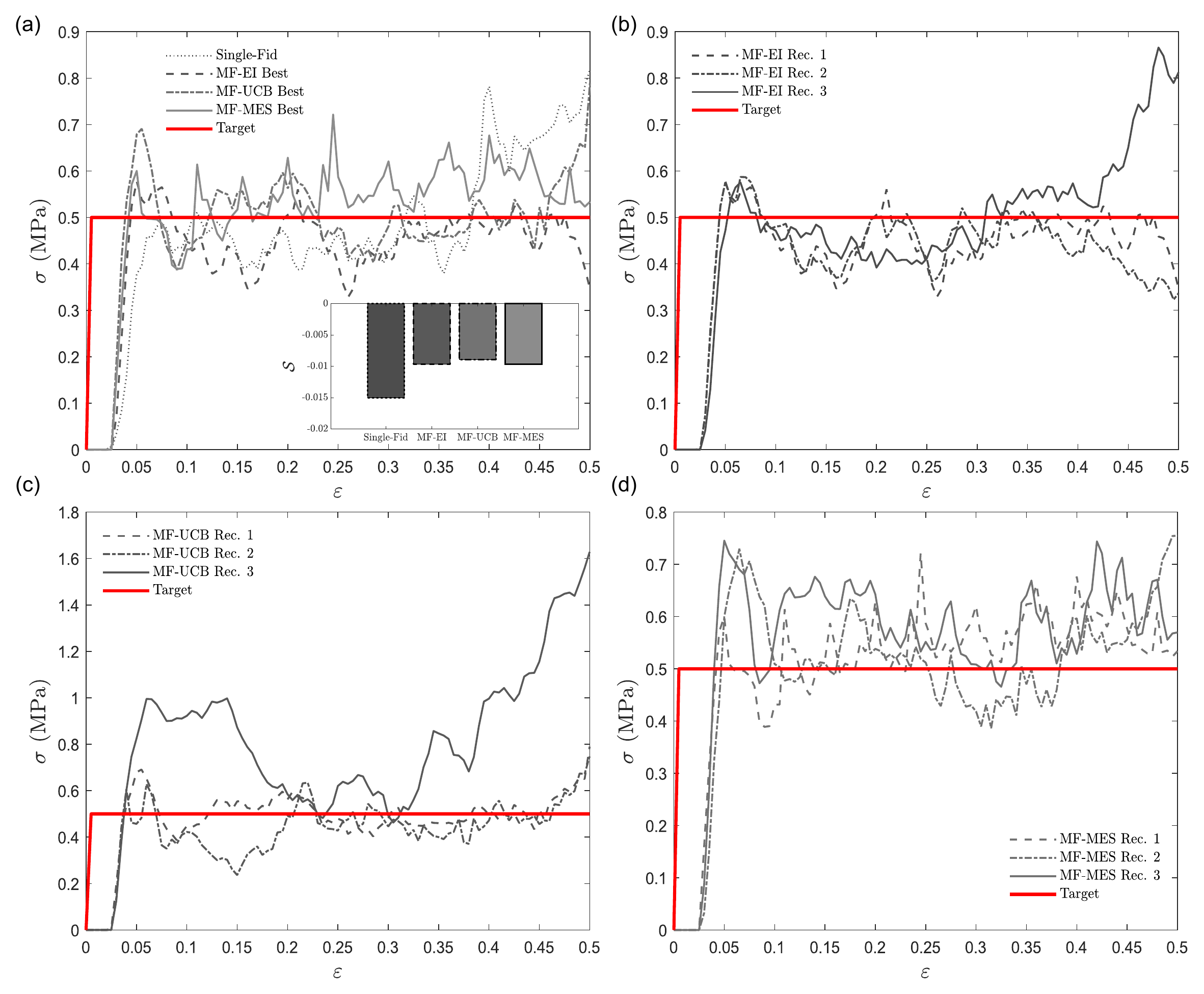}
    \caption{Target 4: (a) Results with the highest similarity score from each method. (b-d) Comparison between the target response and the top three input configurations, based on similarity score, obtained using (b) MF-EI, (c) MF-UCB, and (d) MF-MES.}
    \label{fig:target_4_combined_results}
\end{figure}

\begin{figure}[]
    \centering
    \includegraphics[width=0.9\textwidth]{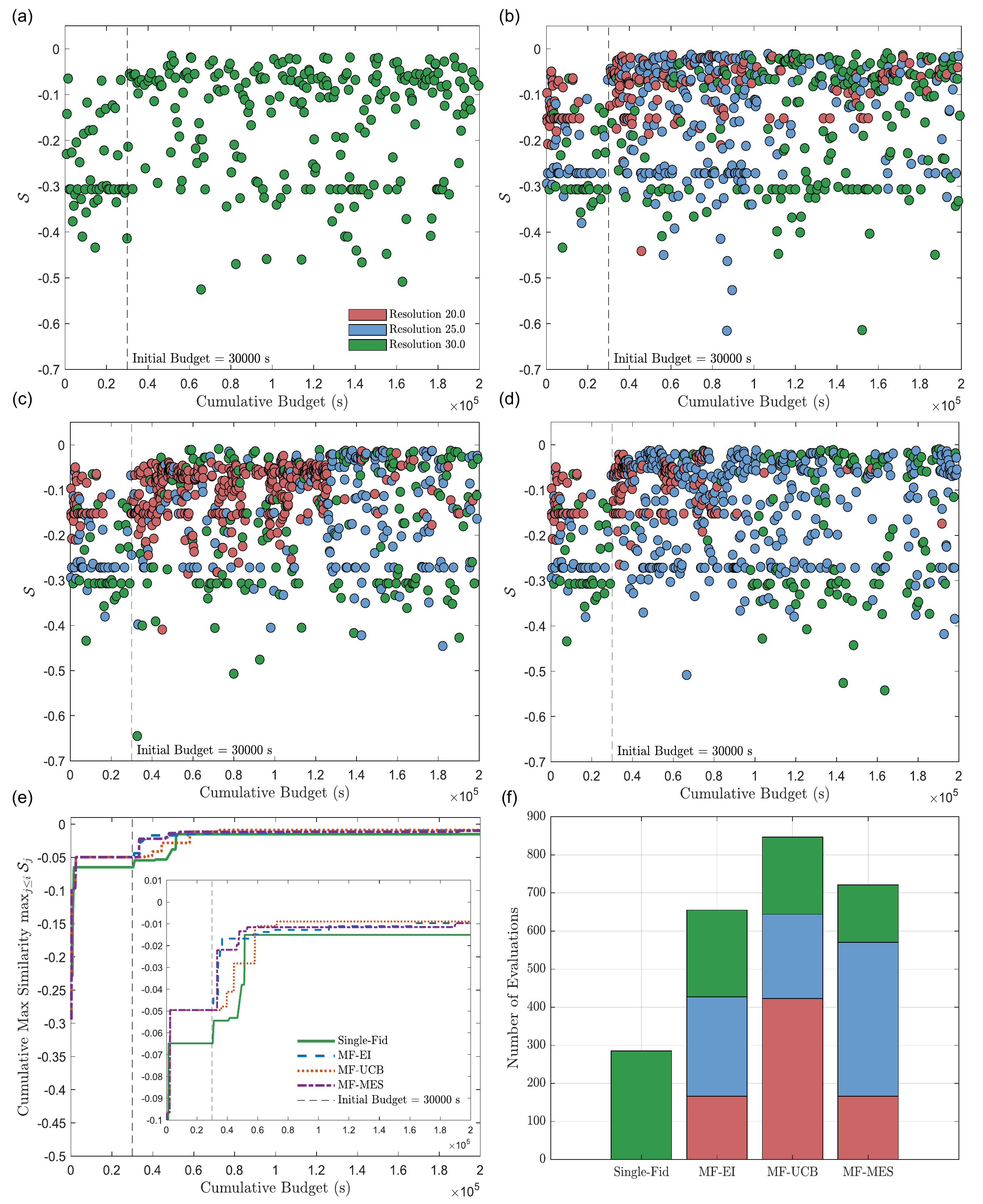}
    \caption[Target 4: Evolution of similarity scores and fidelity usage across four optimisation methods.]{Target 4: Similarity score evolution and fidelity usage for four optimisation methods. (a–d) $\mathcal{S}$ vs.~cumulative computational budget for (a) Single-fidelity, (b) MF-EI, (c) MF-UCB, and (d) MF-MES, with marker colours indicating evaluation fidelity. (e) Cumulative maximum $\mathcal{S}$ comparison under budget constraints, with inset magnifying the overlap region. (f) Stacked bar chart of evaluation counts per fidelity, showing resource allocation across methods.}
    \label{fig:target_4_optimisation_history}
\end{figure}

\section{{Linking design parameters to deformation and failure mechanisms}}
\label{appendix:Linking design parameters to deformation and failure mechanisms}
For Target 4, on average, structures that best replicate the mechanical response of an ideal energy-absorber are obtained with parameters around $\theta_1 \approx 26^{\circ}$; by symmetry in the design space, equivalent structures can also arise at $\theta_2 \approx 26^{\circ}$, with $\theta_3 \approx 0^{\circ}$. These parameter combinations correspond to anisotropic architectures, where the topology is preferentially aligned with the loading direction. They share similarities with the configurations required to attain target responses 1 and 3, but differ in their more intricate geometrical features, which promote progressive, stepwise failure. The resulting geometry is illustrated in the last two columns of Fig.~\ref{fig:ideal_energy_absorber_images}(a).

\begin{figure}[]
    \centering
    \includegraphics[width=1\textwidth]{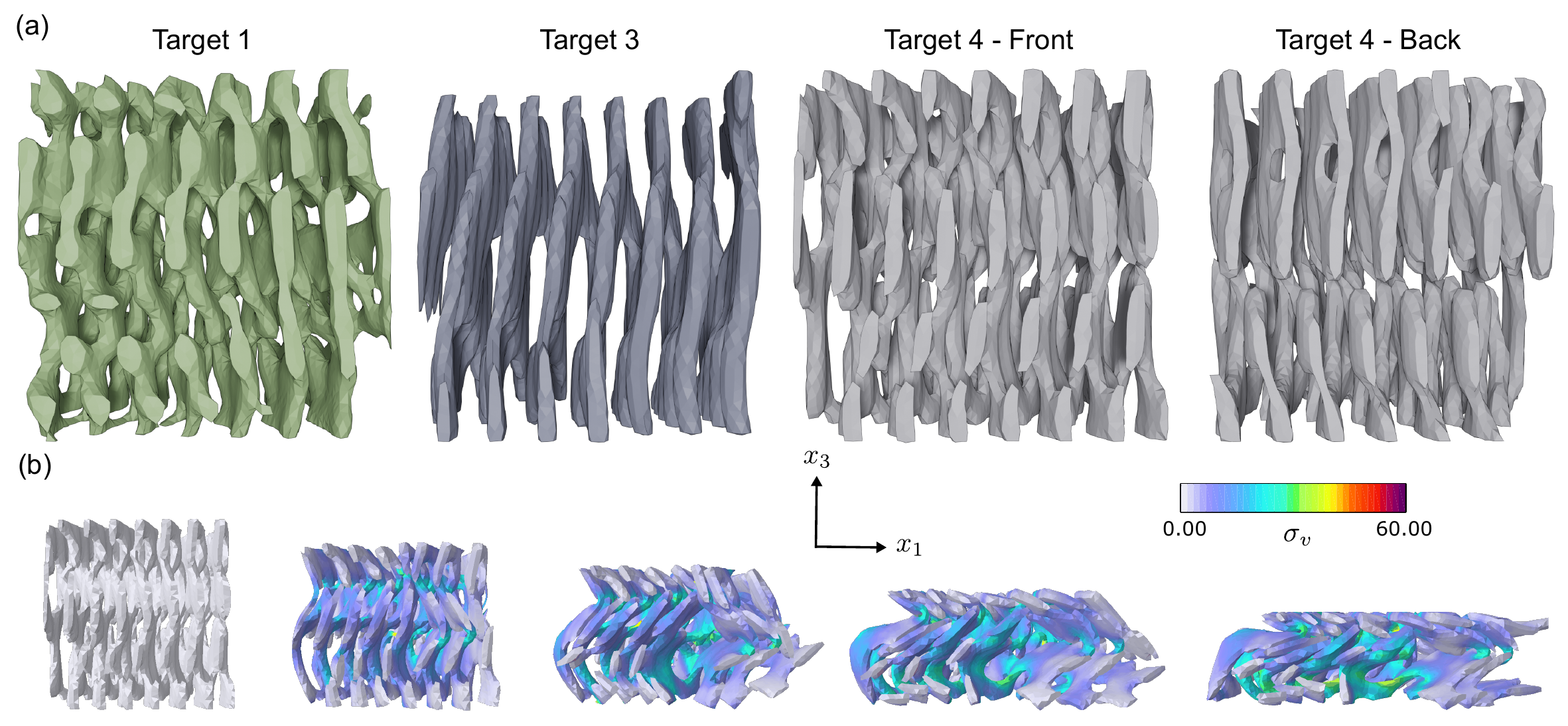}
    \caption[Illustrations of ideal energy absorbing structure generated using target mechanical response 4.]{(a) Three structures corresponding to Target responses 1, 3, and 4. Structure for Target 4 was generated using the parameter set $(\theta_1, \theta_2, \theta_3) = (24.16, 6.07, 0.017)$, which achieved the highest similarity score with the MF-UCB method. (b) Contour plots illustrating the von Mises stress distribution at different loading stages, corresponding to applied displacement values of 0, 5, 10, 15, and 20 mm, respectively.}
    \label{fig:ideal_energy_absorber_images}
\end{figure}

Target 1 (First column of Fig.~\ref{fig:ideal_energy_absorber_images}(a)), generated with $\theta_{1,2,3} = (1.69, 30.44, 6.49)$, produces vertically repeating, interconnected layers. Here, the strong interconnectivity and column thickness reinforce the structure, making it effective at absorbing energy. However, this reinforcement leads to early densification, limiting the plateau region. Target 3, defined by $\theta_{1,2,3} = (15, 5, 0)$, shown in the second column of Fig.~\ref{fig:ideal_energy_absorber_images}(a), also produces a vertically-layered structure, but with thinner and less interconnected layers. Under compression, these slender layers readily buckle once the critical load is exceeded, a process promoted by the curvature of the geometry, which resembles sinusoidal waves. As a result, this architecture is suboptimal for energy absorption, as it collapses too readily.

Target 4 lies between these two cases, with the highest similarity score obtained using the parameters $(\theta_1, \theta_2, \theta_3) = (24.16, 6.07, 0.017)$ from the MF-UCB method, which results in a mechanical response that closely resembles the ideal energy absorber. Generating structures using these parameters yields thin, highly interconnected vertical layers with multiple repeating segments, combining the stabilising benefits of interconnectivity from Target 1 with the slender, buckling-prone features of Target 3. The architecture consists of interlinked slender columns that form leaf-shaped patterns, in contrast to the long, slender layers in Target 3. The layers themselves display a wavy pattern characterised by alternating sinusoidal and cosinusoidal undulations with varying wavelengths, typically shorter at the top and longer towards the bottom. Through the wave-like geometry, the columns bend outwards to form arch-shaped protrusions that split into two prong-like extensions. This graded arrangement ensures that buckling initiates in a distributed and sequential manner, promoting a progressive, stepwise collapse. It is important to note that this structure lacks horizontal reinforcements. Its stiffness arises primarily from the geometry of the columns, especially the curvature, which is a characteristic of a bending-dominated architecture, a hallmark of structures optimal for energy absorption. Similar to the wave-like columns in the Target 3 geometry, those in Target 4 arise from higher-frequency undulations. The increased wave count provides increased support, allowing additional stabilisation under compression, as the more frequent curved segments contribute more material aligned with the loading direction, thereby discouraging premature buckling. 

The rear view (Last column of Fig.~\ref{fig:ideal_energy_absorber_images}(a)) highlights that, upon buckling, the structure can sustain a nearly constant load as each layer interlocks with the gaps of its neighbours. The discontinuities between the vertical layers are apparent; these gaps enable localised deformation and lower the overall stiffness. Moreover, the alternating layers and voids visible from the back suggest that, after a certain level of deformation, specific regions progressively engage, leading to a stepwise deformation sequence. This claw-clip-like arrangement enhances the stability of the buckling process and substantially improves the structure's capacity for energy absorption.

Fig.~\ref{fig:ideal_energy_absorber_images}(b) illustrates the geometry under loading at different displacement intervals. The contour plots highlight a progressive and controlled collapse, marked by stress localisation across multiple regions rather than a single failure zone. Localised thinning and the absence of reinforcement in specific areas drive this behaviour. In particular, the upper and mid-sections of the structure act as stress concentration zones, initiating stiffness degradation. The combined effect of reduced reinforcement and the inherent curvature of the slender beams enables the structure to fold in predetermined regions. This folding mechanism ensures the maintenance of a relatively constant load level, which is a critical characteristic for an ideal energy-absorbing structure.

Nonetheless, it is important to note that while experimentally calibrated fracture energy may alter the recovered parameters for an ideal energy-absorbing structure, the underlying deformation mechanisms, which are progressive bending, buckling, and stepwise collapse, remain unchanged. Moreover, the bending-dominated nature of the structure inherently limits catastrophic damage, making it robust even under variations in post-yield behaviour.

\section{Extension to graded spinodoids}
\label{appendix:Extension to Graded Spinodoids}
The method of generating spinodoids can be extended to produce spatially varying spinodoids. These structures are generated by interpolating the Gaussian random fields (GRFs) used to create individual spinodoid structures. In this approach, three base spinodoids, A, B, and C, are first generated, each defined by a set of input parameters $\Theta_A = (\rho_A, \lambda_A ,\theta_{1A}, \theta_{2A}, \theta_{3A})$, with analogous definitions for structures B and C. The GRF wave number is fixed at $15\pi$. Each base spinodoid is assigned a central location within the cubic domain: structure C at the bottom, B at the centre, and A at the top. The contribution of each spinodoid at any spatial point is then determined using a Gaussian weighting function based on the squared distance from its centre, enabling smooth transitions between the structures. These weights are obtained using

\begin{equation*}
    w_i(\textbf{x}) = \frac{\exp(-\kappa||\textbf{x}-\textbf{c}_i||^2)}{\sum_j\exp(-\kappa||\textbf{x}-\textbf{c}_j||^2} 
\end{equation*}

where $w_i$ is the normalised weight of spinodoid $i$ at position $\textbf{x}$, $\textbf{c}_i$ is the spinodoid's centre, and $\kappa$ controls the sharpness of the transition, fixed to a value of 10. Higher $\kappa$ values produce rapid transitions between structures, whereas lower values result in smoother gradients. It should be noted that the grading direction is dictated by the centre locations.

The final graded spinodoid is then constructed by interpolating the Gaussian random fields $\varphi$ of the base spinodoids according to their weights

\begin{equation*}
    \varphi_{\text{graded}}(\textbf{x}) = \sum_iw_i(\textbf{x})(\varphi_i(\textbf{x}) - \phi_i)
\end{equation*}

where $\phi_i$ is the level set of spinodoid $i$. This procedure ensures a continuous and smooth transition in both density and conical angle from the bottom to the top of the domain. This results in a design space described by 12 input parameters given $\lambda_{A,B,C} = 15\pi$.

Using this approach, two graded structures, one with varying density and the other with varying conical angles, were fabricated and experimentally validated. Fig.~\ref{fig:graded_structure_validation} presents the results, showing good agreement between the experimental measurements and FEM-simulated stress-strain curves, obtained using the same FEM model employed for the simple spinodoids.

\begin{figure}[]
    \centering
    \includegraphics[width=1\textwidth]{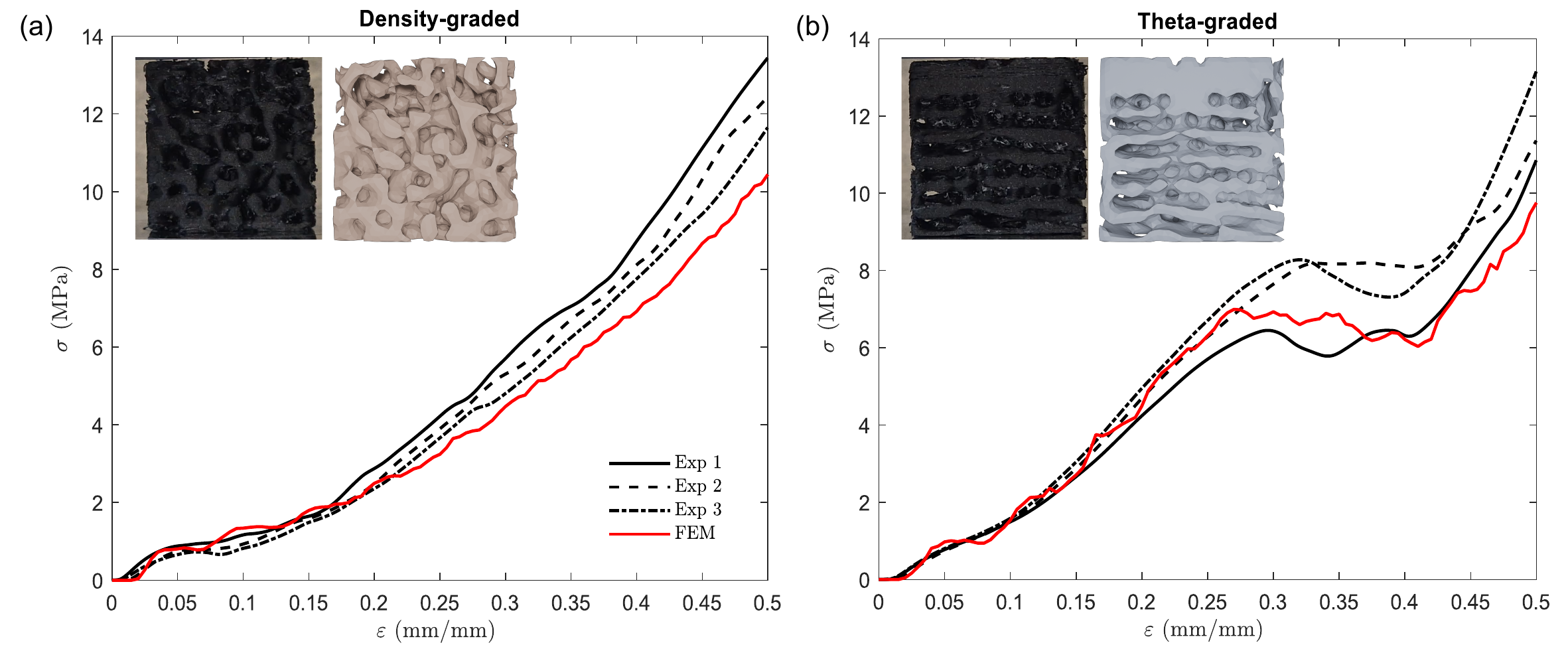}
    \caption[Experimental validation of graded spinodoid topologies.]{Comparison between FEM simulation and experiment results for graded spinodoids, which were graded using two methods: (a) density and (b) conical angles. For (a), $\theta_{1,2,3} = (0, 0, 90)$ were used, with density increasing from 0.3 to 0.45 to 0.6, along the $x_3$ direction from bottom to top. In method (b), the density was held constant at 0.5, while $\theta_{1,2,3}$ varied from $(0, 0, 30)$ to $(20,20,20)$ to $(15,15,0)$ along the same direction. The insets display the 3D printed structure and its corresponding generated model, respectively.}
    \label{fig:graded_structure_validation}
\end{figure}

\end{document}